\documentclass[aip,amsmath,amssymb,reprint]{revtex4-2}

\usepackage{graphicx}
\usepackage{dcolumn}
\usepackage{bm}
\usepackage{mathptmx}
\usepackage{etoolbox}

\makeatletter
\def\@email#1#2{%
 \endgroup
 \patchcmd{\titleblock@produce}
  {\frontmatter@RRAPformat}
  {\frontmatter@RRAPformat{\produce@RRAP{*#1\href{mailto:#2}{#2}}}\frontmatter@RRAPformat}
  {}{}
}%
\makeatother

\usepackage{amsfonts}
\usepackage{physics}
\usepackage{comment}
\usepackage{subcaption}
\usepackage{color}
\usepackage[colorlinks=true,linkcolor=blue,citecolor=red]{hyperref}
\makeatletter
\long\def\@makecaption#1#2{%
  \par
  \begingroup
    \small
    \setlength{\parindent}{0pt}%
    \rightskip=0pt
    \leftskip=0pt
    \parfillskip=0pt
    \spaceskip=0pt
    \xspaceskip=0pt
    \textbf{#1.} #2\par
  \endgroup
}
\makeatother

\def\ve{\varepsilon}
\def\pa{\partial\Omega}
\def\T{{\mathcal T}}

\def\x{\bm{x}}

\def\pa{{\partial \Omega}}
\def\ve{\varepsilon}

\def\T{\mathcal{T}}

\def\X{\mathbf{X}}
\def\N{\mathcal{N}}

\def\md{\mathrm{d}}
\def\llp{\left(}
\def\rrp{\right)}
\def\lla{\left\langle}
\def\rra{\right\rangle}
\def\llv{\left\vert}
\def\rrv{\right\vert}
\def\Var{\mathrm{Var}}

\begin{document}


\title{Correlation between the first-reaction time and the acquired boundary local time}

\author{Yilin Ye} 
\author{Denis~S.~Grebenkov}
\email{denis.grebenkov@polytechnique.edu}
\email{yilin.ye@polytechnique.edu}
\affiliation{Laboratoire de Physique de la Mati\`{e}re Condens\'{e}e (UMR 7643) \\
CNRS -- Ecole Polytechnique, Institut Polytechnique de Paris, 91120 Palaiseau, France
}
\date{\today}

\begin{abstract}
We investigate the statistical correlation between the first-reaction
time of a diffusing particle and its boundary local time accumulated
until the reaction event.  Since the reaction event occurs after
multiple encounters of the particle with a partially reactive
boundary, the boundary local time as a proxy for the number of such
encounters is not independent of, but intrinsically linked to, the
first-reaction time.  We propose a universal theoretical framework to
derive their joint probability density and, in particular, the
correlation coefficient.  To illustrate the dependence of these
correlations on the boundary reactivity and shape, we obtain explicit
analytical solutions for several basic domains.  The analytical
results are complemented by Monte Carlo simulations, which we employ
to examine the role of interior obstacles on correlations in
disordered media.  Applications of these statistical results in
chemical physics are discussed. 
\end{abstract}

\maketitle

\section{Introduction}

The stochastic motion of particles and resulting encounters with their
partners lie at the heart of numerous physical, chemical, and
biological processes \cite{Benichou14, schuss2015brownian, Oshanin,
Grebenkov23c, grebenkov2024target}.  For diffusive particles inside a
bounded domain $\Omega$ with Dirichlet boundary condition, the
statistical characterization of these encounters through the
distribution of first-passage times (FPTs) has been a central theme in
statistical physics for decades \cite{Redner, krapivsky2010kinetic,
schuss2015brownian, Metzler, grebenkov2024target, bray2013persistence,
Masoliver, Oshanin, Dagdug, Meyer11, Singer06, Holcman14,
Mattos12,Grebenkov23c}, where single or multiple targets can be either
inside $\Omega$, or distributed on the boundary $\pa$.  In this
classical framework, a target is considered perfectly absorbing: the
trajectory of a particle terminates upon the first encounter with the
target. The FPT to such a target has been extensively analyzed,
yielding profound insights into diffusion-limited kinetics
\cite{Smoluchowski18, rice1985diffusion}.
However, the first arrival of the particle on the reactive boundary
does not necessarily achieve a successful absorption or reaction
event.  In most realistic scenarios, targets are not perfect sinks
\cite{Grebenkov23c, Guerin21, Guerin23}.  Examples range from trapping
of reactants on catalytic surfaces \cite{bond1986heterogeneous,
Sapoval94, Yuste13} and binding of ligands to cellular receptors
\cite{Berg77, lauffenburger1996receptors} to the search of proteins
for specific DNA sites \cite{halford2004site,
benichou2011facilitated}.  Microscopically, a molecule may interact
with a reactive site multiple times due to insufficient overlapping
between atom orbitals to form chemical bonds, thus being reflected
back into the bulk after each non-reactive collision \cite{Collins49,
Shoup82}.  Such reflected Brownian dynamics result in a random number
of encounters with reactive sites before the successful absorption or
reaction event occurs at the end.  Therefore, the FPT can be expanded
beyond the idealized scenario of a perfectly absorbing boundary to a
more general concept of the first-reaction time (FRT) on a partially
reactive boundary.  In the most common setting, the propensity of the
surface to react is characterized by a constant reactivity $\kappa$,
while the ratio $q = \kappa / D$ quantifies the balance between
diffusive and reactive fluxes on the boundary \cite{Sapoval94,
Piazza22}.  The limit $q \to \infty$ recovers the perfect absorption
such that the FRT is reduced to the FPT, while $q=0$ describes a
perfectly reflecting, inert surface and the FRT turns out to be
infinite.

Two particles with identical FRTs may have explored the partially
reactive boundary to vastly different extents.  A complete description
of the reaction pathway thus requires a proper measure of this
boundary exploration. This quantitative role is precisely fulfilled by
the boundary local time (BLT) \cite{Levy,Ito,Freidlin}, which can be
expressed as the rescaled number of encounters with the reactive
boundary region $\Gamma$:
\begin{equation}  \label{eq:ellt_def2}
\ell_t = \lim\limits_{\ve\to 0} \ve \, \N_t^{(\ve)} ,
\end{equation} 
where $\N_t^{(\ve)}$ is the number of crossings of the thin layer
$\Gamma_\ve = \{ \x\in\Omega: \llv \x - \Gamma \rrv < \ve \}$ of width
$\ve$ near the reactive region $\Gamma$ up to time $t$.
Alternatively, the BLT $\ell_t$ quantifies the residence time in this
boundary layer up to time $t$,
\begin{equation}  \label{eq:ellt_def1}
\ell_t = \lim\limits_{\ve \to 0} \frac{D}{\ve} \int\limits_0^t \md t' \, \Theta(\ve - |\X_{t'} - \Gamma|),
\end{equation} 
where $D$ is the diffusion coefficient, $|\x - \Gamma|$ is the
Euclidean distance between a point $\x$ and the reactive region
$\Gamma$, and $\Theta(z)$ is the Heaviside step function: $\Theta(z) =
1$ for $z > 0$ and $0$ otherwise.  

The BLT is used to construct the encounter-based approach to
diffusion-controlled reactions \cite{Grebenkov20}, allowing one to
describe different surface reaction mechanisms by imposing suitable
stopping conditions
\cite{Grebenkov21,Bressloff22,Grebenkov22,Bressloff22d,Benkhadaj22,Grebenkov23b}.
In this paper, we consider a broad class of surface reactions,
in which the reaction event occurs after a sufficient number of
encounters of the particle with the reactive region $\Gamma$.  More
formally, we introduce an independent random variable $\hat{\ell}$
with some probability density $\psi(\ell)$ and postulate that the
first-reaction time on $\Gamma$ is the first instance when the BLT
$\ell_t$ exceeds the threshold $\hat{\ell}$:
\begin{equation} 
\tau = \inf \{ t > 0 : \ell_t > \hat\ell \} . \label{eq:tauinf} 
\end{equation} 
In this way, the BLT $\ell_t$ incorporates the diffusive dynamics in
$\Omega$, whereas the threshold $\hat{\ell}$ describes the chemical
kinetics on the reactive region.  For instance, the most common
scenario of a partially reactive boundary with a constant reactivity
$\kappa = qD$ corresponds to the exponential threshold with the
density $\psi(\ell) = q e^{-q\ell}$ (see \cite{Grebenkov20} for
details).

This probabilistic formulation explicitly connects the macroscopic
parameter $q$ to a microscopic stopping condition for the BLT.  Still,
the acquired BLT does not uniquely constrain the FRT. Even though two
diffusive particles starting from the same location may acquire the
same BLT, their FRTs can be substantially different.  In other words,
one cannot infer the FRT from the BLT, nor vice versa; in fact, these
random variables provide complementary information on the reaction
event.  This raises a fundamental and yet unexplored question: How
statistically correlated are the first-reaction time $\tau$ and the
boundary local time $\ell_\tau$ accumulated until the instance of
reaction on $\Gamma$?  In particular, the correlation coefficient $C_q
= \mathrm{Corr}(\tau, \ell_\tau)$ that measures the degree to which
reaction time predicts boundary exploration (and vice versa), is
crucial for deeper understanding of the diffusion-reaction pathways.

The paper is organized as follows. In Sec. \ref{sec:theo}, we propose
a general approach to derive the joint probability distribution of the
FRT $\tau$ and the BLT $\ell_\tau$ and, in particular, their
correlation coefficient. Section \ref{sec:num} describes Monte Carlo
simulations, their validation for simple symmetric domains, and
numerical results for disordered media with inert obstacles. We
discuss further extensions and conclude the paper in
Sec. \ref{sec:discon}.

\section{Theoretical results}
\label{sec:theo}

Our analysis relies on the encounter-based approach
\cite{Grebenkov20}.  We consider ordinary diffusion in a bounded
domain $\Omega \subset \mathbb{R}^d$ whose smooth boundary $\pa =
\Gamma \bigcup \pa_N$ is split into a partially reactive region
$\Gamma$ and the remaining reflecting part $\pa_N$.  For the boundary
local time $\ell_t$ on $\Gamma$, we introduce the first-crossing time
(FCT) $\mathcal{T}_\ell = \inf \{ t> 0: \ell_t > \ell \}$ of a
\textit{fixed} threshold $\ell$.  The probability density
$U(\ell,t|\x_0)$ of the random variable $\mathcal{T}_\ell$ depends on
the starting point $\x_0 \in \Omega$, the threshold $\ell$, and the
geometry of the confining domain $\Omega$. In particular, its Laplace
transform
\begin{equation}
\tilde{U}(\ell,p|\x_0) = \left\langle e^{-p \mathcal{T}_\ell} \right\rangle 
= \int\limits_0^\infty \mathrm{d} t \, e^{-pt} U(\ell,t|\x_0) \quad (p \ge 0) , 
\end{equation}
admits the spectral expansion \cite{Grebenkov20} :
\begin{equation}
\tilde{U}(\ell,p|\x_0) = \sum_{k=0}^\infty e^{-\mu_k^{(p)} \ell} \ V_k^{(p)} (\x_0) \int\limits_\Gamma \mathrm{d} \x \ V_k^{(p)}(\x) \ ,
\label{eq:Uspec}
\end{equation}
where $\mu_k^{(p)}$ and $V_k^{(p)}(\x)$ are the eigenvalues and
eigenfunctions of the generalized Steklov problem:
\begin{subequations}
\begin{align}
(p - D \Delta) V_k^{(p)}(\x) & = 0 \qquad (\x \in \Omega), \\
\partial_n V_k^{(p)} (\x) & = \mu_k^{(p)} V_k^{(p)} \qquad (\x \in \Gamma), \\
\partial_n V_k^{(p)} & = 0 \qquad (\x \in \pa_N) ,
\end{align}
\end{subequations}
where $\partial_n$ is the normal derivative oriented outward from the
domain $\Omega$. The spectrum of this problem is known to be discrete
\cite{Levitin23}, and the eigenpairs $\{ \mu_k^{(p)}, V_k^{(p)} \}$
are enumerated by $k = 0, 1, 2, \ldots$ to order the eigenvalues
$\mu_k^{(p)}$ into an increasing sequence.  The $m$th-order moment of
the FCT can be found from the moment-generating function
$\tilde{U}(\ell,p|\x_0)$:
\begin{equation}
\langle \mathcal{T}^m_\ell \rangle = \int\limits_0^\infty \md t \ t^m \ U(\ell,t|\x_0) 
=  (-1)^m \left( \partial^m_p \tilde{U}(\ell,p|\x_0) \right)_{p=0} .
\end{equation}

Let us now show that the probability density $U(\ell,t|\x_0)$ plays
the key role for understanding correlations between the first-reaction
time $\tau$ and the BLT $\ell_\tau$ prior to the reaction event.
According to Eq. (\ref{eq:tauinf}), the first-reaction time is the FCT
of the random threshold $\hat{\ell}$, i.e., $\tau =
\mathcal{T}_{\hat{\ell}}$  and thus $\ell_\tau =
\hat{\ell}$.  As a consequence, the joint probability density of
$\ell_\tau$ and $\tau$ is
\begin{equation}   \label{eq:Pjoint}
\mathcal{P}(\ell,t|\x_0) = \psi(\ell) \, U(\ell,t|\x_0).
\end{equation}
In fact, once the random threshold $\hat{\ell}$ is selected with the
density $\psi(\ell)$, the first-reaction time is chosen with the
density $U(\ell,t|\x_0)$.  The product in Eq. (\ref{eq:Pjoint}) simply
reflects that the threshold was supposed to be independent of the
diffusive dynamics (and thus of the confining domain); in this way,
the threshold characterizes exclusively the chemical kinetics on the
reactive region.  This conceptually important extension of
$U(\ell,t|\x_0)$ was unnoticed in previous works, and presents one of
the main results of this paper.  Note that the (marginal)
probability density of $\tau$ is obtained by integrating
$\mathcal{P}(\ell,t|\x_0)$ over $\ell$, whereas the (marginal)
probability density of $\ell_\tau$ is obtained by integrating over
$t$, which expectedly yields $\psi(\ell)$ due to the normalization of
$U(\ell,t|\x_0)$:
\begin{equation}  \label{eq:Unorm}
\int\limits_0^\infty \mathrm{d} t \ U(\ell,t|\x_0) = 1.
\end{equation}

In the following, we focus on the most common scenario of a partially
reactive region with a constant reactivity $\kappa = qD$, for which
$\psi(\ell) = q e^{-q\ell}$.  In this case, the probability density of
the first-reaction time, denoted as $H_q(t|\x_0)$, turns out to be
proportional to the Laplace transform of $U(\ell,t|\x_0)$ with respect
to $\ell$:
\begin{equation}
H_q(t|\x_0) = \int\limits_0^\infty \mathrm{d} \ell \ {\mathcal P}(\ell,t|\x_0) 
= \int\limits_0^\infty \mathrm{d} \ell \ q e^{-q\ell} \ U(\ell,t|\x_0) .
\end{equation}
For any integer $m>0$, one can express the $m$th-order moment of the
first-reaction time as:
\begin{align} 
\langle \tau^m \rangle &= \int\limits_0^\infty  \mathrm{d} \ell \int\limits_0^\infty  \mathrm{d} t \ t^m \ {\mathcal P}(\ell,t|\x_0)  
= \int\limits_0^\infty \md \ell \ q e^{-q \ell} \langle \mathcal{T}^m_\ell \rangle . 
\label{eq:tauT}
\end{align}
An alternative calculation can be realized by the following recurrence
relations for $m \geqslant 1$:
\begin{subequations}
\begin{align}
D \Delta \langle \tau^{m} \rangle = - m \langle \tau^{m-1} \rangle &\qquad (\x \in \Omega), \label{eq:taum10a} \\ \label{eq:taum10b}
\partial_n \langle \tau^{m} \rangle + q \langle \tau^{m} \rangle = 0 &\qquad (\x \in \Gamma) , \\
\partial_n \langle \tau^m \rangle = 0 &\qquad (\x \in \pa_N) , 
\end{align}
\label{eq:taumrecurr} 
\end{subequations}
with $\langle \tau^0 \rangle = 1$.   As discussed in
\cite{Grebenkov20}, the Robin boundary condition (\ref{eq:taum10a}) is
tightly related to the chosen exponential form of $\psi(\ell)$; in
particular, another probability law for the threshold $\hat{\ell}$
would result in a different boundary condition.

In turn, the $m$th-order moment of the acquired boundary local time
$\ell_\tau$ is universal and independent of the domain $\Omega$:
\begin{equation}
\langle \ell_\tau^m \rangle = \int\limits_0^\infty \mathrm{d} \ell \ \ell^m \int\limits_0^\infty \mathrm{d} t \ {\mathcal P}(\ell,t|\x_0) 
= \frac{m!}{q^m} ,
\label{eq:ellmoment}
\end{equation}
where we used again the normalization (\ref{eq:Unorm}) of the
probability density $U(\ell,t|\x_0)$.  This is simply the $m$th-order
moment of the chosen exponentially distributed threshold $\hat{\ell}$.
In particular, one gets the variance of $\ell_\tau$ as:
\begin{equation} 
\mathrm{Var}\{ \ell_\tau \} = \frac{1}{q^2} . \label{eq:ellvar} 
\end{equation}  
We also compute the crossing average as:
\begin{equation}
\langle \tau \ell_\tau \rangle = \int\limits_0^\infty \mathrm{d} \ell \ \ell 
\int\limits_0^\infty \mathrm{d} t \ t \ qe^{-q\ell} \ U(\ell,t|\x_0) 
= \int\limits_0^\infty \mathrm{d} \ell \ \ell \ q e^{-q \ell} \langle \mathcal{T}_\ell \rangle .
\label{eq:crossing}
\end{equation}
Combining these results, we determine the correlation coefficient
between $\tau$ and $\ell_\tau$:
\begin{equation}
C_q = \mathrm{Corr} \{ \tau, \ell_\tau \} = \frac{\langle \tau \ell_\tau \rangle 
- \langle \tau \rangle \langle \ell_\tau \rangle }{\sqrt{\mathrm{Var}\{\tau\} \mathrm{Var}\{\ell_\tau\} } } .
\label{eq:cq}
\end{equation}
Integrating Eq. \eqref{eq:crossing} by parts, one can also obtain the
following representation:
\begin{equation}
\langle \tau \ell_\tau \rangle - \langle \tau \rangle \langle \ell_\tau \rangle 
= \int\limits_0^\infty \md \ell \ \ell e^{-q \ell} \partial_\ell \langle \mathcal{T}_\ell \rangle .
\label{eq:crosspt}
\end{equation}
As $ \langle \mathcal{T}_\ell \rangle$ is an increasing function of
$\ell$, the correlation is always positive.

\subsection{Asymptotic behavior}
\label{sec:asymp}

Let us examine the asymptotic behavior of the correlation coefficient
$C_q$.

In the limit of high reactivity ($q\to \infty$), both $\langle
\tau\rangle$ and $\Var\{\tau\}$ approach their limits $\langle
\tau_\infty\rangle$ and $\Var\{\tau_\infty\}$ corresponding to the
first-passage time $\tau_\infty$ to a perfectly absorbing boundary
region $\Gamma$. Moreover, the Laplace-transform relation
\begin{equation}  \label{eq:tau_Tell}
\langle \tau \rangle = \int\limits_0^\infty \md\ell \, q e^{-q\ell} \, \langle \T_\ell\rangle  
\end{equation}
(a particular case of Eq. (\ref{eq:tauT}) for $m=1$) reveals that the
large-$q$ behavior of $\langle \tau \rangle$ corresponds to the
small-$\ell$ behavior of $\langle \T_\ell\rangle$. As $\ell\to 0$, the
mean first-crossing time $\langle \T_\ell\rangle$ also approaches
$\langle \tau_\infty\rangle$: in fact, the first crossing of the
threshold $\ell = 0$ occurs at the first passage to the boundary.  As
a consequence, Eq. (\ref{eq:crossing}) implies that $\langle \tau
\ell_\tau \rangle \approx \langle \tau_\infty\rangle/q$.  Substituting
these expressions into Eq. (\ref{eq:cq}), we conclude that $C_q \to 0$
as $q\to \infty$. In other words, the first-reaction time $\tau$ and
the acquired boundary local time $\ell_\tau$ become uncorrelated in
this limit. This is not surprising: for a perfectly reactive boundary,
one has $\ell_\tau = 0$, whereas $\tau$ obeys the probability density
$H_\infty(t|\x_0)$, independently of $\ell_\tau$.

In the opposite limit of low reactivity ($q\to 0$), the mean FRT is
dominated by slow reaction kinetics because the particle has to
perform multiple reaction attempts and thus has enough time to diffuse
through the bounded domain $\Omega$.  Integrating
Eq. (\ref{eq:taum10a}) with $m = 1$ over $\Omega$, one gets
\begin{equation} 
|\Omega| = \int\limits_{\Omega} \md\x_0 \, (-D \Delta \langle \tau\rangle) 
= qD \int\limits_{\Gamma} \md\x_0 \, \langle \tau\rangle .
\label{eq:Omega_tau}
\end{equation}
As the starting point does not matter in the leading order in this
regime, one has $\langle \tau\rangle \approx$ const, and thus
\begin{equation}
\langle \tau\rangle \approx \frac{|\Omega|}{qD |\Gamma|} \quad (q\to 0).
\label{eq:taumean}
\end{equation}
In the same vein, one can integrate Eq. (\ref{eq:taum10a}) for the
second moment ($m=2$) to get
\begin{equation}
\int\limits_{\Omega} \md\x_0 \, 2\langle \tau\rangle = qD \int\limits_{\Gamma} \langle \tau^2\rangle ,
\label{eq:taum2}
\end{equation}
from which the same argument yields
\begin{equation}
\langle \tau^2\rangle \approx 2 \langle \tau\rangle \frac{|\Omega|}{qD |\Gamma|} 
\approx 2 \biggl(\frac{|\Omega|}{qD |\Gamma|}\biggr)^2 \quad (q\to 0),
\end{equation}
and thus $\sqrt{\Var\{\tau\}} \approx |\Omega|/(qD |\Gamma|)$.
Finally, we employ again Eq. (\ref{eq:tau_Tell}) to show that
\begin{equation}
\langle \T_\ell \rangle \approx \frac{|\Omega| \ell}{D|\Gamma|} 
\label{eq:Tlmean}
\end{equation}
in the large-$\ell$ limit that corresponds to $q\to 0$.  As a
consequence, we get
\begin{equation}
\langle \tau \ell_\tau \rangle \approx \frac{|\Omega|}{D|\Gamma|} \int\limits_0^\infty \md\ell \, \ell^2 \, q e^{-q\ell} 
= \frac{2|\Omega|}{q^2 D|\Gamma|} \,.
\end{equation}
Substituting these expressions into Eq. (\ref{eq:cq}), we conclude
that $C_q \to 1$ as $q\to 0$.  This result is also not surprising: in
the reaction-limited regime, the particle undergoes numerous reaction
attempts, and the first-reaction time can be seen as the sum of random
durations of diffusive motions in the bulk between successive
encounters with the boundary; this sum is asymptotically proportional
to the number of attempts, i.e., $\tau \sim \ell_\tau$, so that the
two variables become fully correlated.

\subsection{Special case for the starting point}
\label{sec:starting_Gamma}

It is instructive to consider the special case when the starting point
is uniformly distributed over the reactive boundary region $\Gamma$.
Substituting Eq. (\ref{eq:tau_Tell}) into Eq. (\ref{eq:Omega_tau}), we
get
\begin{equation}
\frac{|\Omega|}{q^2 D |\Gamma|} = \int\limits_0^\infty \md\ell \, e^{-q\ell} \, \overline{\langle \T_\ell\rangle},
\label{eq:eq24}
\end{equation}
where
\begin{equation}
\overline{\langle \T_\ell\rangle} = \frac{1}{|\Gamma|} \int\limits_{\Gamma} \md\x_0 \, \langle \T_\ell \rangle 
\end{equation}
is the average (denoted by overline) of the mean FCT over the starting
point uniformly distributed on $\Gamma$. Inverting the Laplace
transform (\ref{eq:eq24}), we find
\begin{equation}
\overline{\langle \T_\ell\rangle} = \frac{|\Omega| \ell}{D |\Gamma|} \,,
\end{equation}
which is exact and valid for any confining domain. As a consequence,
we obtain from Eq. (\ref{eq:crosspt}):
\begin{align}
 \frac{1}{|\Gamma|}\int\limits_{\Gamma} \md\x_0 \bigl[\langle \tau \ell_\tau\rangle 
- \langle \tau \rangle \langle \ell_\tau\rangle\bigr] 
=& \int\limits_0^\infty \md\ell \,\, \ell \,\, e^{-q\ell} \,\, \partial_\ell \biggl(\frac{|\Omega| \ell}{D |\Gamma|}\biggr) \nonumber \\
=& \frac{|\Omega|}{q^2 D |\Gamma|} ,
\end{align}
and the correlation coefficient in this case becomes
\begin{equation}
\overline{C_q} = \frac{|\Omega|}{q D |\Gamma| \sqrt{\overline{\mathrm{Var} \{ \tau \} }} } , 
\end{equation}
where $\overline{\mathrm{Var} \{ \tau \} }= \overline{\langle \tau^2
\rangle} - \left( \overline{\langle \tau \rangle} \right)^2 $ is the
variance of the FRT with the uniform starting point on $\Gamma$.  The
boundary-averaged mean $\overline{\langle \tau \rangle}$ is given by
Eq. (\ref{eq:Omega_tau}):
\begin{equation}
\overline{\langle \tau \rangle} = \frac{1}{|\Gamma|} \int\limits_{\Gamma} \mathrm{d} \x_0 \langle \tau \rangle = \frac{|\Omega|}{q D |\Gamma|} ,
\end{equation}
whereas the second moment can be expressed from Eq. (\ref{eq:taum2}): 
\begin{equation}
\overline{\langle \tau^2 \rangle} = \frac{1}{|\Gamma|} \int\limits_{\Gamma} \mathrm{d} \x_0 \langle \tau^2 \rangle 
= \frac{2|\Omega|}{q D |\Gamma|} {T}_q ,
\end{equation}
where ${T}_q = \frac{1}{|\Omega|} \int\limits_\Omega \mathrm{d} \x_0 \langle \tau \rangle$. 
We get thus 
\begin{equation}
\overline{C_q} = \frac{1}{\sqrt{ 2 q D {T}_q |\Gamma| / |\Omega| - 1} } ,
\end{equation} 
i.e., this correlation coefficient is expressed in terms of the
volume-averaged mean FRT ${T}_q$.  This exact relation exemplifies the
transition between two asymptotic limits discussed in
Sec. \ref{sec:asymp}: as $q\to0$, one can use again
Eq. (\ref{eq:taumean}), written as $T_q \simeq |\Omega| / (q D
|\Gamma|)$, to see that $\overline{C_q} \to 1$; in turn, as
$q\to\infty$, $T_q$ approaches the volume-averaged mean first-passage
time $T_\infty$, and $\overline{C_q}$ vanishes. We will show below
that the predicted decay $\overline{C_q} \propto q^{-1/2}$ is specific
to the choice of the starting point on the reactive region $\Gamma$.

\subsection{Example of a ball}

To illustrate the above results, we consider diffusion in a ball
$\Omega = \{ \x \in \mathbb{R}^3 : |\x| < R \}$ of radius $R$, with a
partially reactive boundary $\pa = \Gamma$ (here $\pa_N =
\emptyset$), for which the Steklov eigenvalues and eigenfunctions are
known explicitly (see, e.g., \cite{Grebenkov20,
grebenkov2020surface}). The rotational symmetry implies that only the
rotationally invariant eigenfunction $V_0^{(p)}(\x) =
\frac{1}{\sqrt{|\Gamma|}} \frac{i_0(\alpha |\x|)}{i_0(\alpha R)}$
contributes to Eq. (\ref{eq:Uspec}), yielding
\begin{align}
\tilde{U}(\ell,p|\x_0) &= e^{-\mu_0^{(p)} \ell} V_0^{(p)}(\x_0) \int\limits_\Gamma \md\x \ V_0^{(p)}(\x) \nonumber \\
&= e^{-\mu_0^{(p)} \ell} \frac{i_0(\alpha |\x_0|)}{i_0(\alpha R)} ,
\end{align}
where $\alpha = \sqrt{p/D}$, $\mu_0^{(p)} = \alpha i_0^\prime(\alpha
R) / i_0(\alpha R)$, $i_0(z) = \sinh(z)/z$ is the modified spherical
Bessel function of the first kind and zero order, and $i_0^\prime(z) =
\mathrm{d} i_0(z) / \mathrm{d} z$.
For a starting point $|\x_0| = r $ inside the ball, one retrieves the
Laplace transform of the FRT probability density:
\begin{align}
\tilde{H}_q (p|r) &= \int\limits_0^\infty \mathrm{d} \ell \ q e^{-q\ell} \ \tilde{U}(\ell,p|\x_0) \nonumber \\ 
&= \frac{q i_0(\alpha r)}{\alpha i_0^\prime(\alpha R) + q i_0(\alpha R)} ,
\end{align}
from which one gets 
\begin{subequations}
\begin{align}
\lla \tau \rra (r) =& \frac{R^2 - r^2}{6D} + \frac{R}{3Dq} , \label{eq:tau1_sphere} \\
\lla \tau^2 \rra (r) =& \frac{3 r^4-10 r^2 R^2+7 R^4}{180 D^2} \nonumber \\ 
& + \frac{R \left(7 R^2-5 r^2\right)}{45 D^2 q} + \frac{2 R^2}{9 D^2 q^2}\\
\mathrm{Var} \{ \tau \} (r) =& \frac{R^4 - r^4}{90 D^2} + \frac{10+4Rq}{90D^2q^2} R^2 .
\end{align}
\end{subequations}
We have thus
\begin{equation}
\lla \mathcal{T}_\ell \rra (r) = \frac{R^2 - r^2}{6D} + \frac{R \ell}{3D} ,
\end{equation}
while Eq. (\ref{eq:crossing}) implies 
\begin{align}
\lla \tau \ell_\tau \rra (r) &= \int\limits_0^\infty \md \ell \ \ell \ qe^{-q\ell} 
\llp \frac{R^2 - r^2}{6D} + \frac{R \ell}{3D} \rrp \nonumber \\
&= \frac{R^2 - r^2}{6Dq} + \frac{2R}{3Dq^2} ,
\end{align}
and thus
\begin{equation}
\lla \tau \ell_\tau \rra - \lla \tau \rra \lla \ell_\tau \rra = \frac{R}{3Dq^2} . 
\label{eq:cm_sphere}
\end{equation}
Combining these results with Eq. (\ref{eq:ellvar}), we determine the
correlation coefficient between $\tau$ and $\ell_\tau$ from
Eq. (\ref{eq:cq}):
\begin{equation}
C_q (r) = \frac{1}{\sqrt{1 + 2qR/5 + (qR)^2(1-(r/R)^4) / 10}} . \label{eq:Cq_sphere_r}
\end{equation}
One can easily check the asymptotic behavior described in
Sec. \ref{sec:asymp}: $C_q(r) \to 1$ as $q \to 0$ and $C_q(r) \to 0$
as $q \to \infty$. Moreover, we quantify now how fast or slow these
limits are achieved; in particular, as $q\to\infty$, one gets $C_q(r)
\propto q^{-1}$ for any $r<R$ and $C_q(R) \propto q^{-1/2}$, as we saw
in Sec. \ref{sec:starting_Gamma}.   A similar computation for the disk
of radius $R$ yields (see Appendix \ref{app:disk}):
\begin{equation}
C_q (r) = \frac{1}{\sqrt{1 + qR/2 + (qR)^2 \left(1-(r/R)^4\right) / 8}} . 
\end{equation}

In Appendix \ref{app:sphere}, we also derive the correlation
coefficient for the case when the starting point is uniformly
distributed inside the ball (we use the symbol $\circ$ to refer to
this setting):
\begin{equation}
C_q (\circ) = \frac{1}{\sqrt{1 + 2qR/5 + 13(qR)^2/175}} .
\label{eq:Cq_sphere_circ}
\end{equation}
One can see the same $2qR/5$ term in both Eqs. (\ref{eq:Cq_sphere_r})
and (\ref{eq:Cq_sphere_circ}), implying the independence of $C_q$ on
the starting point up to order $q R$ in the limit $qR \ll 1$.
Similarly, the correlation coefficient for uniform starting points in
the disk of radius $R$ is (see Appendix \ref{app:disk})
\begin{equation}
C_q (\circ) = \frac{1}{\sqrt{1 + q R / 2 + 5 (q R)^2 / 48 }} .
\label{eq:Cq_disk_circ}
\end{equation}

\subsection{Concentric spherical shell $(d \ge 2)$}
\label{sec:shells}

Following the approach above, we now analyze diffusion inside a
concentric spherical shell of radii $R$ and $L$. The inner boundary
$\Gamma$ is partially reactive with the parameter $q>0$, while the
outer boundary $\pa_N$ is inert.  For this symmetric domain, the
relation $\lla \tau \ell_\tau \rra - \lla \tau \rra \lla \ell_\tau
\rra = \frac{|\Omega|}{|\Gamma| Dq^2}$ holds.  We present detailed
calculations for the circular annulus ($d=2$) and the spherical shell
($d=3$) in Appendices \ref{app:annulus} and \ref{app:shell},
respectively.

It is instructive to inspect the correlation coefficient in the
small-target limit $R \ll L$ and moderate reactivity.  Expanding the
correlation coefficient in powers of the small parameters $qR$ and
$R/L$, we obtain in two dimensions (see Appendix \ref{app:annulus}):
\begin{widetext}
\begin{subequations}
\begin{align}
C_q(r) =& 1 - qR \left( \log \left(\frac{L}{R} \right) - \frac{3}{4}\right) \nonumber +\frac{(q R)^2}{32} \left[72 \log \left(\frac{R}{L}\right) + 48 \log^2 \left(\frac{R}{L}\right) \right. \\ &\left. + 8 \log \left(\frac{r}{R}\right) \left( 3 + 2\log \left(\frac{r R}{L^2}\right) \right)  + 27 - 16 \left(\frac{r}{L}\right)^2 + 2 \left(\frac{r}{L}\right)^4\right] + O\left(R^3\right) , \\
C_q(\circ) =& 1 - qR \left( \log \left(\frac{L}{R} \right) - \frac{3}{4}\right) +\frac{(q R)^2}{12} \left[6 \log \left(\frac{L}{R}\right) \left(2 \log \left(\frac{L}{R}\right)-3\right)+5\right] + O\left(R^3\right) . 
\end{align}
\label{eqs:Cq2d}
\end{subequations}
\end{widetext}
Curiously, the leading-order behavior of the correlation coefficient
$C_q(r)$ is independent of the starting point $\x_0$ (note that $r =
|\x_0|$ appears starting from the second-order term).  Similarly, for
diffusive particles near a tiny spherical target inside a large ball
$(d=3)$, the initial position does not affect the correlation
coefficient up to order $R^3$ (see Appendix \ref{app:shell}):
\begin{widetext}
\begin{subequations}
\begin{align}
C_q(r) =& 1 - q R + (qR)^2 \left( 1 + \frac{9}{5q L} \right) - (qR)^3  \left( 1 + \frac{18}{5q L} \right) \nonumber \\ &+\frac{q R^4}{100 L^6 r^2} \left(50 L^6 \left(2 q^3 r^2+q\right)+180 L^5 q r (3 q r-1)+486 L^4 q r^2-100 L^3 r^2 (q r+3)+5 q r^6\right) + O\left(R^5\right) , \\
C_q(\circ) =& 1-q R + (qR)^2 \left( 1 + \frac{9}{5q L} \right) - (qR)^3  \left( 1 + \frac{18}{5q L} \right) +\frac{q R^4 (L q (35 L q (5 L q+27)+459)-525)}{175 L^3}+O\left(R^5\right) . 
\end{align}
\label{eqs:Cq3d}
\end{subequations}
\end{widetext}

These asymptotic results clearly show that the FRT and the BLT are
strongly correlated in the small-target limit with moderate reactivity
$q$ (such that $qR \ll 1$). We stress that the reactivity itself does
not need to be small: one needs $ q \ll q_{\mathrm{max}}$, with the
maximal value $q_{\mathrm{max}} = 1/ R$ being large as $R \to 0$.
While Eqs. (\ref{eqs:Cq2d}, \ref{eqs:Cq3d}) were derived for a
particular geometry of a spherical shell, we expect that the above
conclusion remains valid in general.

\newpage
\section{Numerical results}
\label{sec:num}

In the previous section, we have shown that the probability
density $U(\ell,t|\x_0)$ of the first-crossing time $\T_\ell$
determines the joint probability density $\mathcal{P}(\ell,t|\x_0)$
and controls correlations between the first-reaction time $\tau$ and
the acquired boundary local time $\ell_\tau$.  In particular, the
theoretical framework based on the generalized Steklov problem gives
access to the moment-generating function $\tilde{U}(\ell,p|\x_0)$ of
$\T_\ell$ via its spectral expansion.  We illustrated its efficient
application for rotation-invariant domains.  In general, however, the
Steklov eigenfunctions are not known explicitly, and one has to resort
to numerical techniques such as a finite-element method
\cite{Chaigneau24,Grebenkov25a} or a boundary-integral method
\cite{Nigam25}.  In the particular case when the reactive region $\Gamma$
is small, one can employ matched asymptotic techniques to get explicit
but approximate results
\cite{Bressloff22,Grebenkov22,Grebenkov25b,Grebenkov26}.

In this section, we choose a different strategy and employ Monte Carlo
simulations to provide complementary insights into correlations
between the FRT and the BLT. Sections \ref{sec:MC} and
\ref{sec:valid} present the simulation algorithm and its validation,
whereas Sections \ref{sec:2dregular},
\ref{sec:2drandom}, \ref{sec:3dregular} summarize our main numerical
results for restricted diffusion in disordered media.

\begin{figure}[]
  \begin{subfigure}[b]{0.1\textwidth}
{\includegraphics[trim={2.9cm 0.7cm 2.9cm 0.7cm}, clip, height=2.4cm]{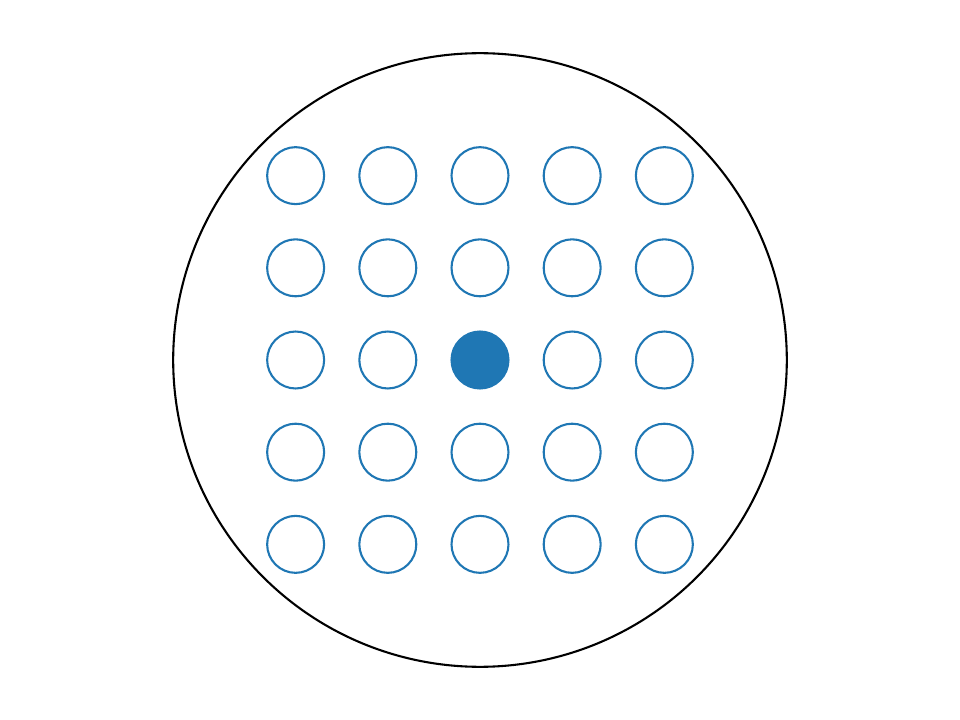} }
  \caption{}  \label{fig:dm5sq}
  \end{subfigure}
\hspace{3.0em} 
  \begin{subfigure}[b]{0.1\textwidth}
{ \includegraphics[trim={1.9cm 0.6cm 1.9cm 0.6cm}, clip, height=2.4cm]{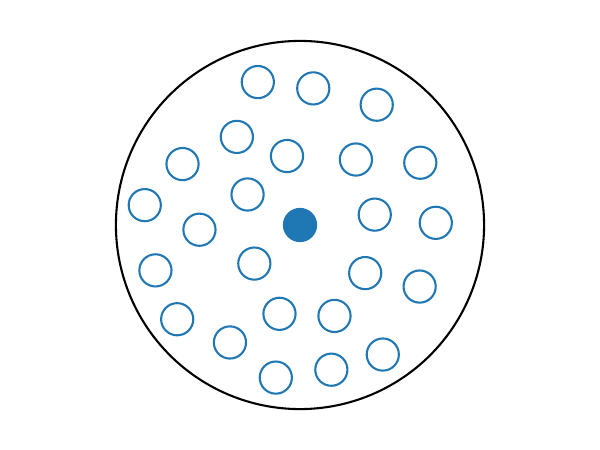} }
  \caption{}  \label{fig:dm5r}
  \end{subfigure}
  \hspace{3.0em} 
\begin{subfigure}[b]{0.1\textwidth}
{\includegraphics[trim={2cm 1.8cm 1.8cm 1.6cm}, clip, height=2.4cm]{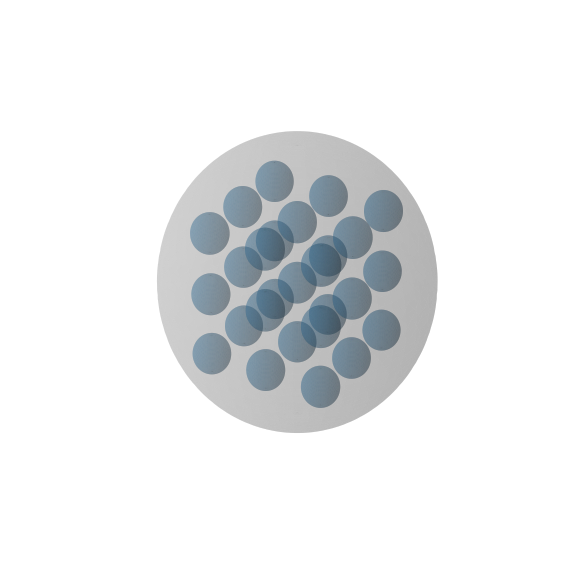} }
\caption{} \label{fig:dm3d3}
\end{subfigure}
\caption{
Computational domains with interior obstacles of the same radius $R$:
the central one (in blue) is the reactive target, while the others are
inert.
\textbf{(a)} 
Regular square-lattice packing of $m \times m$ identical disks inside
the unit disk, with $m=5$.  The shown configuration has $R = 0.618 \,
R_{\mathrm{max}}$, with the maximal radius $R_{\mathrm{max}} =
L/(\sqrt{2}(m-1)+1) \approx 0.15$.
\textbf{(b)} Random packing of 25 disks with the same $R$.
The coordinates of all disks are provided in Appendix
\ref{app:random25}.  
\textbf{(c)} 
Regular cubic-lattice packing of $m \times m \times m$ identical balls
inside the unit ball with $m=3$, and $R = 0.618 \,
R_{\mathrm{max}}$, with the maximal radius $R_{\mathrm{max}} =
L/(\sqrt{3}(m-1)+1) \approx 0.22$.  }
\label{fig:domain_all}
\end{figure}

To investigate the role of geometric disorder, we consider domains
containing interior inert obstacles (Fig. \ref{fig:domain_all}).  We
study three packing configurations:
(a) a regular arrangement of identical disks of radius $R$ inside the
unit disk;
(b) a random packing of identical non-overlapping disks of radius $R$
inside the unit disk;
(c) a regular arrangement of identical balls of radius $R$ inside the
unit ball, with the central ball being reactive.
In all three cases, the central disk (ball) is partially reactive,
whereas all the others are reflecting (inert obstacles).

\subsection{Description of Monte Carlo simulations}
\label{sec:MC}

To numerically investigate the joint statistics of the first-reaction
time $\tau$ and the acquired boundary local time $\ell_\tau$, we
perform extensive Monte Carlo simulations.  For a given domain
$\Omega$, starting points $\x_0$ are sampled from a uniform
distribution inside $\Omega$.  Each random trajectory $\X_t$ with
$\X_{0} = \x_0$ is generated as follows.  The walk-on-spheres (WOS)
method \cite{Muller56} is employed for rapidly generating random
points of a simulated trajectory inside $\Omega$, whereas the BLT
remains unchanged.  This sampling is repeated until the particle
enters a thin boundary layer $\pa_\ve = \{ \x \in \Omega : |\x - \pa|
< \ve \}$ of a prescribed width $\ve$.  In this case, the
escape-from-a-layer (EFL) approach \cite{ye2025escape} is used to
accurately compute the increment of the BLT acquired until the escape
of the particle from that layer. In this way, the most time-consuming
simulation of reflected Brownian motion inside the layer is replaced
by an equivalent escape event.  Note that the BLT is incremented only
in the vicinity of the reactive region $\Gamma$ whereas its increments
near the reflecting part $\pa_N$ are ignored.  After the escape from
$\pa_\ve$, the particle resumes its bulk diffusion simulated by WOS,
and so on.  The simulation of the trajectory is stopped when the
acquired BLT $\ell_t$ exceeds an independently generated threshold
$\hat{\ell}$, obeying the exponential distribution with a given
parameter $q$. This stopping time is recorded as the FRT $\tau$,
whereas the current value of the BLT is recorded as $\ell_\tau$.
After that, the next trajectory is simulated from a newly generated
starting point $\x_0$.  Repeating this procedure $N$ times yields an
empirical joint distribution of the FRT and the acquired BLT. From
this dataset, we compute statistical moments (means, variances) and
the correlation coefficient $C_q(\circ)$ according to
Eq. (\ref{eq:cq}).  We generally use $N = 10^6$ to ensure
statistically robust estimates except for few cases explicitly
specified.  The proximity threshold $\varepsilon$ is chosen
sufficiently small, $\varepsilon = 10^{-3}$, to guarantee that the
error in approximating the FRT and the BLT is negligible.

\subsection{Validation}
\label{sec:valid}

\begin{figure}[t!]
  \centering
  \begin{subfigure}[b]{0.48\textwidth}
\includegraphics[width=0.99\linewidth]{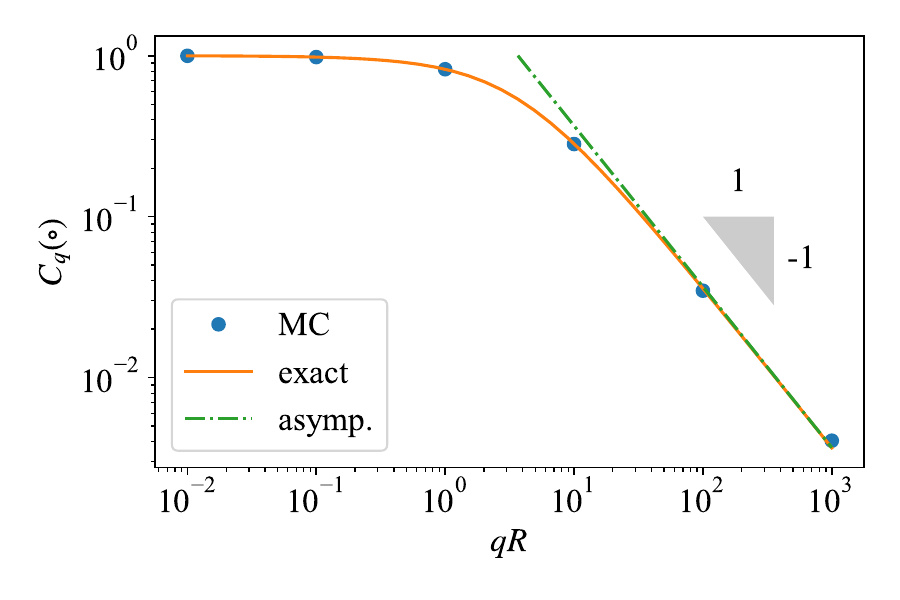}    \caption{Unit ball}
    \label{fig:corrtl_3d_1}
  \end{subfigure}
  \hspace{1.0em} 
  \begin{subfigure}[b]{0.48\textwidth}
\includegraphics[width=0.99\linewidth]{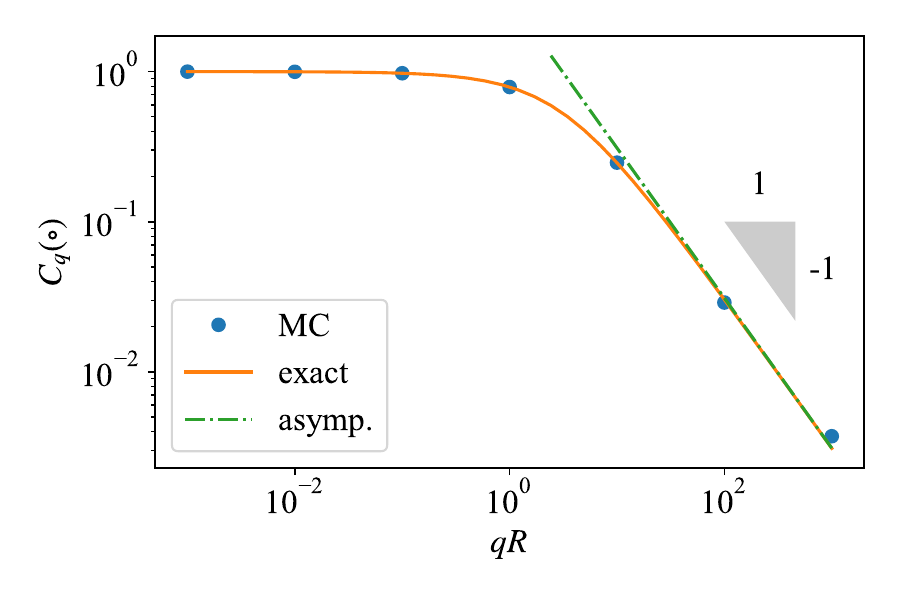}      \caption{Unit disk}
    \label{fig:corrtl_2d_1}
  \end{subfigure}
\caption{
Correlation coefficient $C_q(\circ)$ for \textbf{(a)} the unit ball and
\textbf{(b)} the unit disk ($R=1$).  Comparison between Monte Carlo
results (MC, blue dots) and analytical expressions
(\ref{eq:Cq_sphere_circ}, \ref{eq:Cq_disk_circ}) (orange curves).
Numerical statistics are obtained from $N=10^6$ particles, with
$D=1$. Green dashed lines indicate the asymptotic behaviors
$C_q(\circ) \approx 5\sqrt{7/13}/(qR)$ for $d=3$ and $C_q(\circ)
\approx 4 \sqrt{3/5} / (qR)$ for $d=2$ in the limit $q\to\infty$.  }
\label{fig:corrtl_simple}
\end{figure}

Our Monte Carlo method is validated by comparing the correlation
coefficients $C_q(\circ)$ obtained from simulations with the exact
analytical expressions (\ref{eq:Cq_sphere_circ},
\ref{eq:Cq_disk_circ}) for uniform starting points inside the unit
ball ($d=3$) or the unit disk ($d=2$).  As shown in
Fig. \ref{fig:corrtl_3d_1}, the numerical results (blue dots) are in
excellent agreement with Eq. (\ref{eq:Cq_sphere_circ}) (orange curve)
across a broad range of $q$, from $10^{-2}$ to $10^3$.  In the limit
$q\to0$, one gets $C_q \to 1$, indicating that the FRT $\tau$ and the
BLT $\ell_\tau$ become fully correlated.  As $q$ increases,
$C_q(\circ)$ decreases monotonically.  In the large reactivity limit
$q\to\infty$, a leading-order expansion of
Eq. (\ref{eq:Cq_sphere_circ}) yields the asymptotic behavior
$C_q(\circ) \propto 1/q$.  The same behavior is observed for the unit
disk as shown in Fig. \ref{fig:corrtl_2d_1}, where the simulation data
perfectly match the exact solution given by
Eq. (\ref{eq:Cq_disk_circ}).

\begin{figure}[t]
\centering
\includegraphics[width=0.495\linewidth]{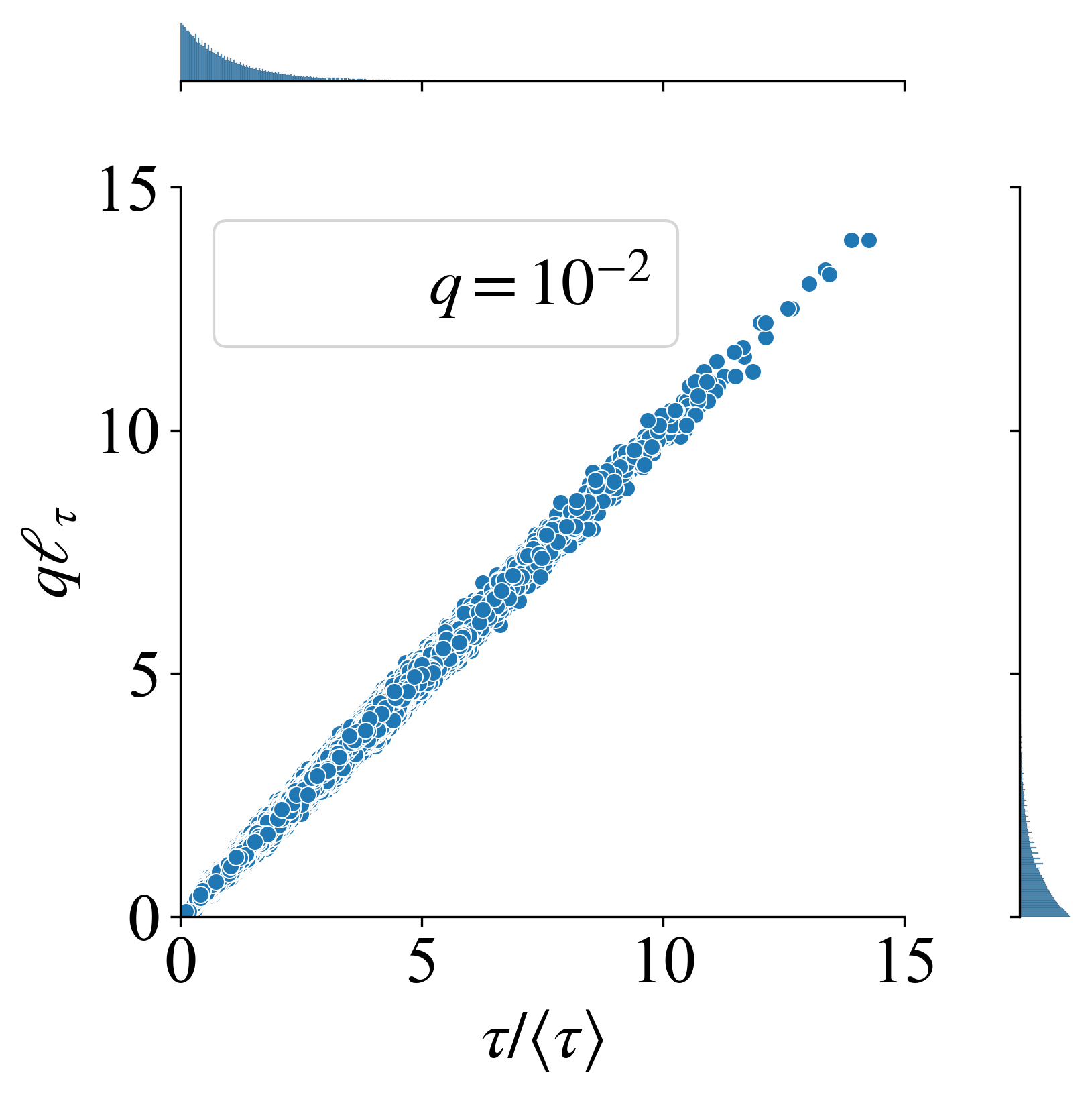}
\includegraphics[width=0.495\linewidth]{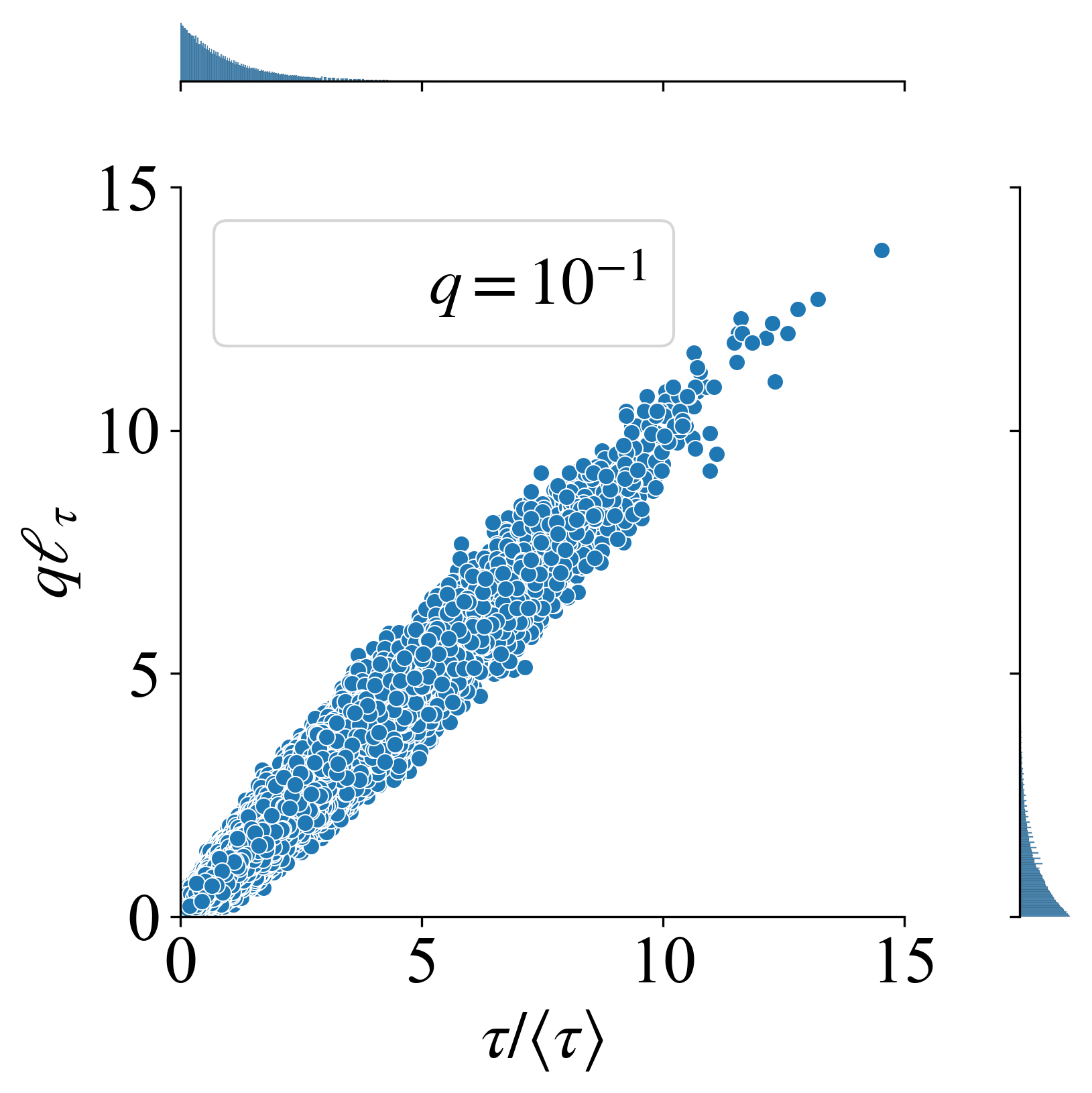} \\
\includegraphics[width=0.495\linewidth]{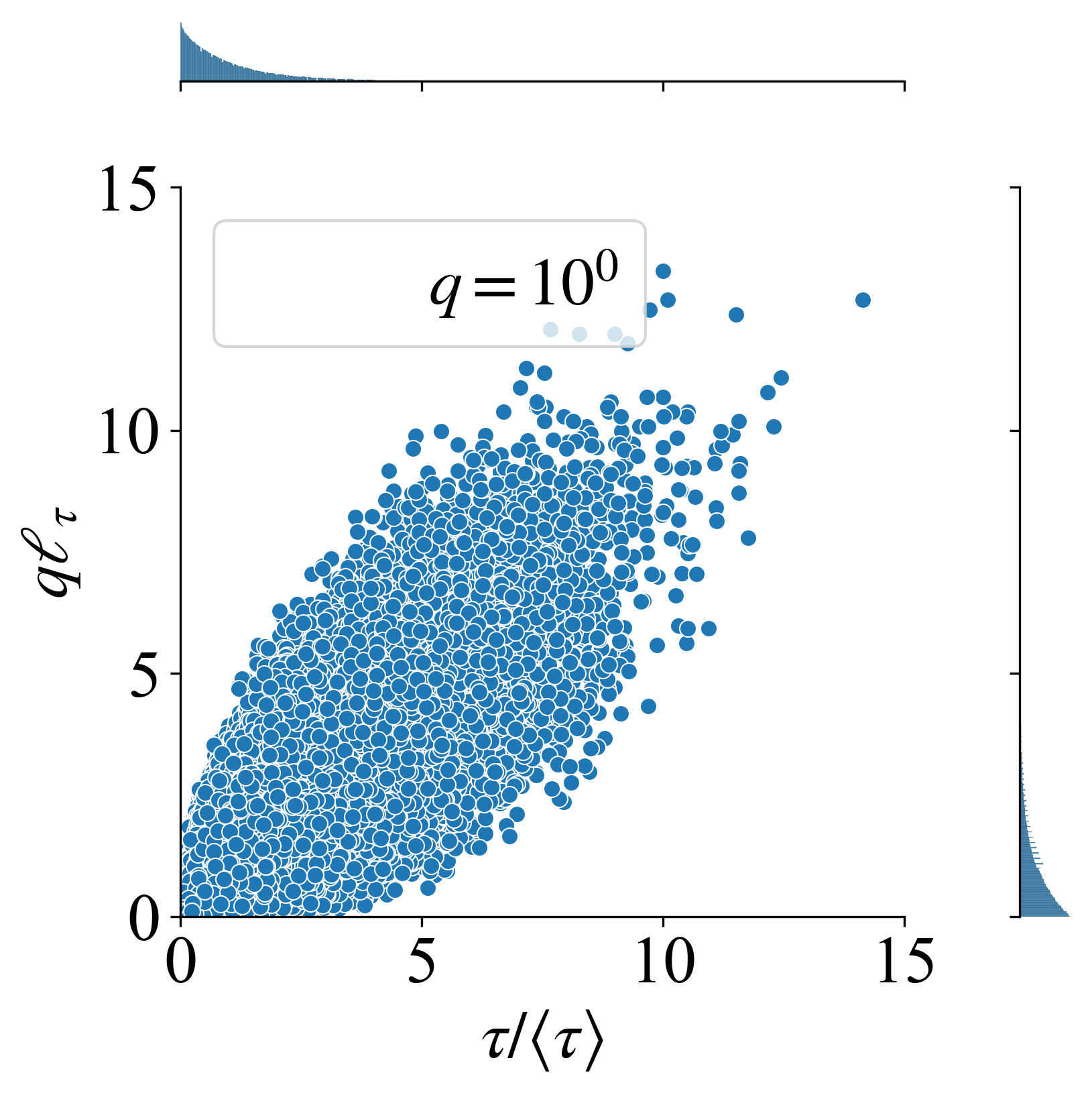}
\includegraphics[width=0.495\linewidth]{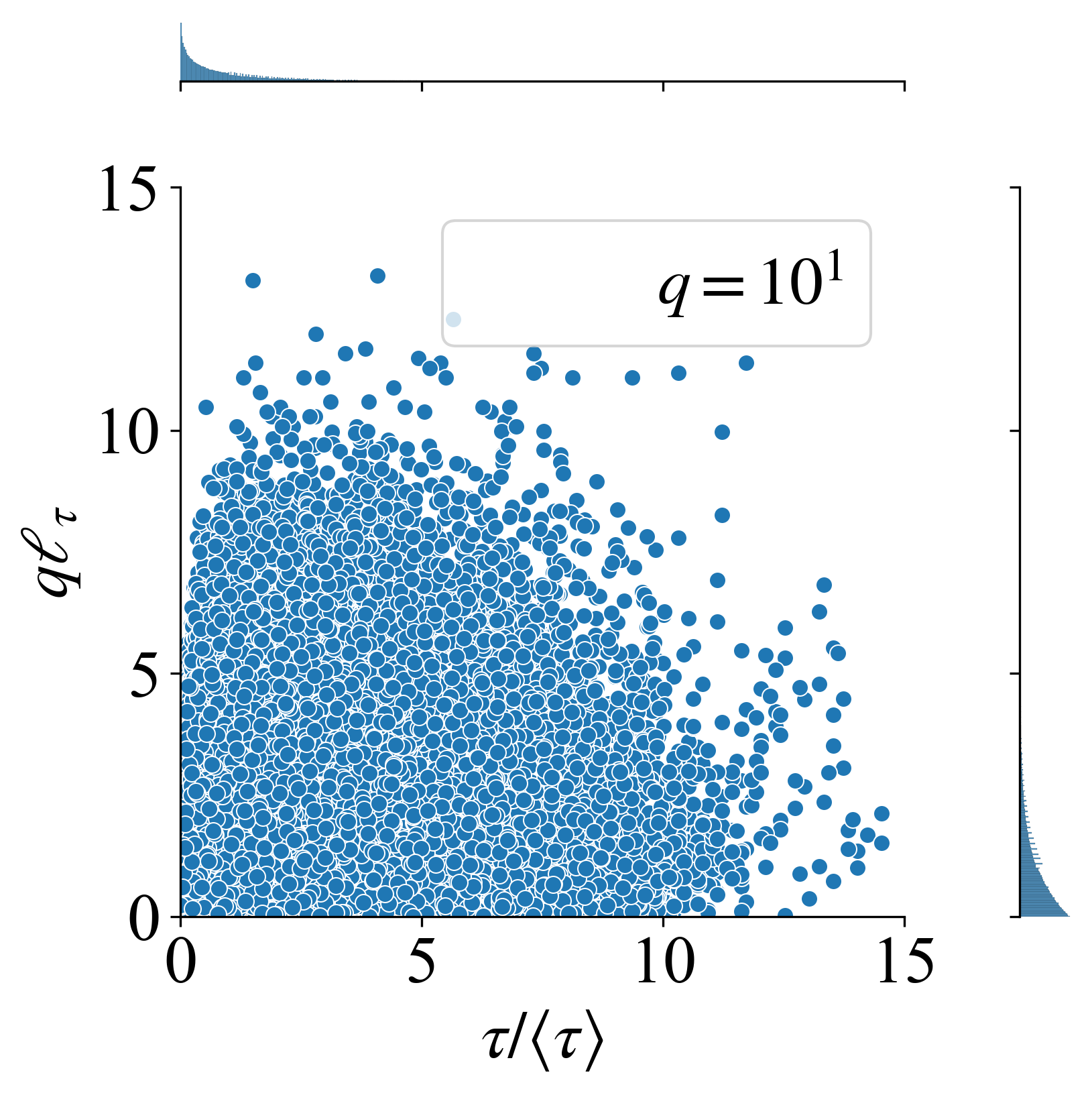}
\caption{
Joint histograms of the FRT $\tau$ and the BLT $\ell_\tau$ for $q =
10^{-2}, 10^{-1}, 10^0, 10^1$ obtained by Monte Carlo simulations with
$N=10^6$ particles, where the starting points are uniform inside the
unit ball. Both $\tau$ and $\ell_\tau$ are normalized by their
respective mean values.  Each blue dot represents a random realization
of the pair $(\tau,\ell_\tau)$, whereas their scatter visualizes
correlations.  The marginal distributions of $\tau /
\langle \tau \rangle$ and $\ell_\tau / \langle \ell_\tau \rangle = q
\ell_\tau$ are shown above and right of each joint plot.  }
\label{fig:corrtl_sphere}
\end{figure} 

Beyond the correlation coefficient itself, the joint histogram of
$\tau$ and $\ell_\tau$ offers a more direct visual representation of
their statistical coupling.  Figure \ref{fig:corrtl_sphere} presents
such joint histograms in the case of uniform starting points inside
the unit ball at different reactivities.  The random variables $\tau$
and $\ell_\tau$ are normalized by their mean values given by
Eq. (\ref{eq:tau1_sphere_unif}) and Eq. (\ref{eq:ellmoment}),
respectively.  For large $q$, the pairs $(\tau, \ell_\tau)$ are widely
scattered, reflecting weak correlations.  In turn, for smaller
reactivity, pairs concentrate along a straight line $\ell_\tau /
\langle \ell_\tau \rangle \approx \tau / \langle \tau \rangle$,
indicating the strong linear relation between $\tau$ and $\ell_\tau$
anticipated from the limit $C_q \to 1$ as $q\to0$.  In addition, we
present the marginal distributions of $\tau$ and $\ell_\tau$ that
exhibit similar exponential decay across all $q$.  One sees that most
random realizations of $(\tau, \ell_\tau)$ are concentrated near the
lower bottom corner of each panel, i.e., in the region where
$\tau/\langle\tau\rangle$ and $\ell_\tau/\langle \ell_\tau\rangle$ are
of the order of $1$ (or smaller).

\subsection{Regularly packed obstacles in 2d}
\label{sec:2dregular}

\begin{figure}[t!]
  \centering
  \begin{subfigure}[b]{0.48\textwidth}
\includegraphics[width=0.99\linewidth]{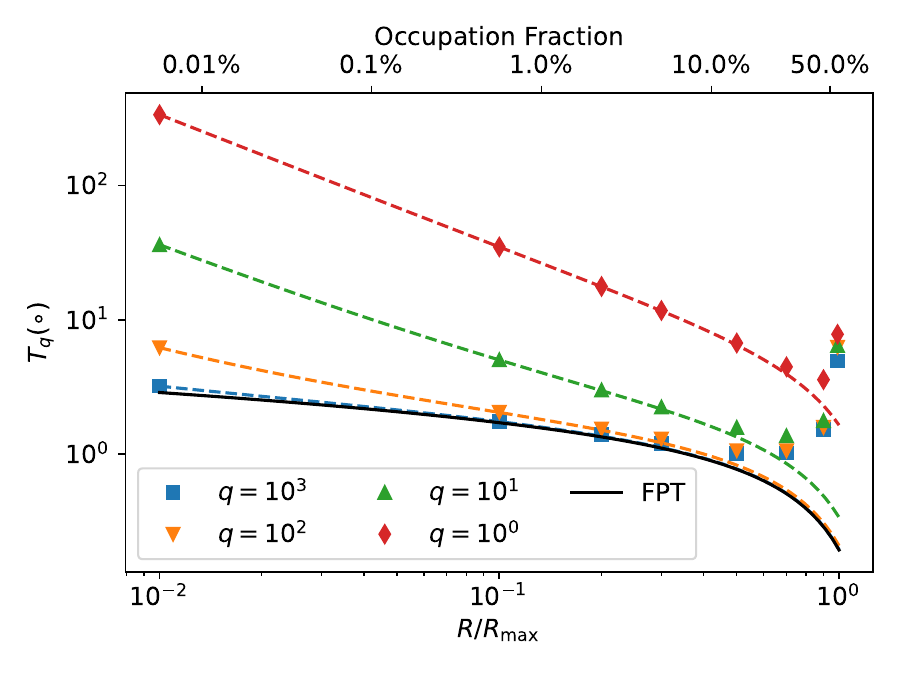}    \caption{Mean FRT}
    \label{fig:T5}
  \end{subfigure}
  \hspace{1.0em} 
  \begin{subfigure}[b]{0.48\textwidth}
\includegraphics[width=0.99\linewidth]{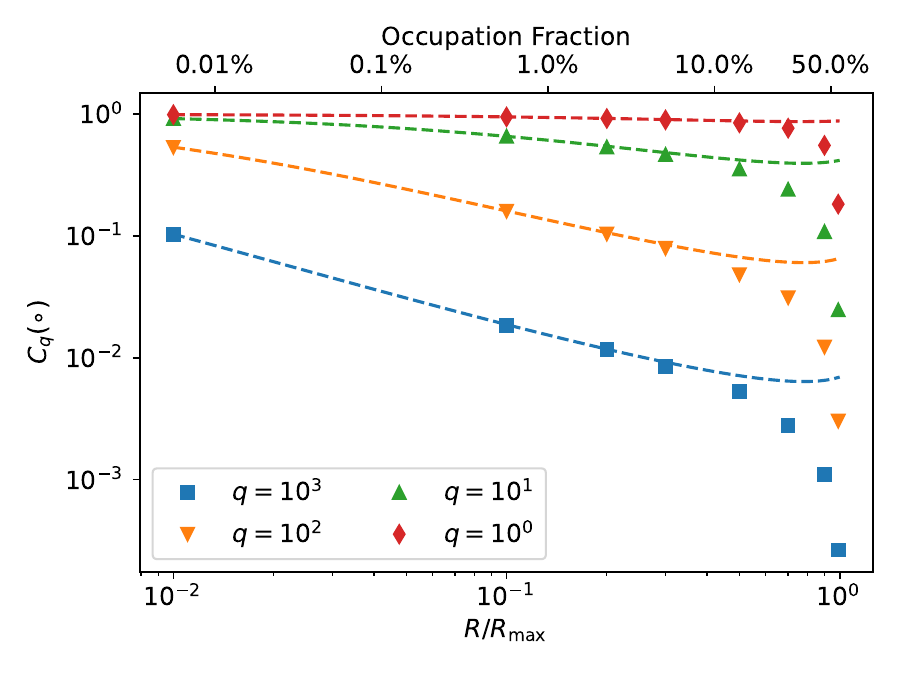}    \caption{Correlation coefficient}
    \label{fig:Cq5}
  \end{subfigure}
\caption{
Mean FRT and correlation coefficient for a regular arrangment of
obstacles (the domain shown in Fig. \ref{fig:dm5sq} with $m = 5$).
Symbols correspond to different reactivities: $q=10^3$ (squares),
$q=10^2$ (downward triangles), $q=10^1$ (upward triangles), and
$q=10^0$ (diamonds).  Numerical results are obtained by Monte Carlo
simulations with $N=10^6$ particles that are initially uniformly
distributed inside the domain(for $q = 10^3$, we used $N = 10^7$ and
$\ve = 10^{-4}$).
\textbf{(a)} Mean FRT as a function of the scaled disk radius
$R/R_{\mathrm{max}}$. The solid black curve represents the mean
first-passage time (i.e., $q=\infty$), and colored dashed curves show
the annulus approximation (\ref{eq:tau1_annulus_circ}).  \textbf{(b)}
Correlation coefficient $C_q(\circ)$ as a function of scaled disk
radius. Colored dashed curves correspond to the annulus approximation
(\ref{eq:Cq_annulus_circ}), in which $L$ is replaced by
$L_{\mathrm{eff}}$ from Eq. (\ref{eq:annulus_app}).  Occupation
fraction of obstacles is shown on the horizontal top axis.}
\label{fig:m5_2d}
\end{figure}

We employ Monte Carlo simulations to inspect the impact of geometric
disorder onto the correlation coefficient.  As illustrated by
Fig. \ref{fig:dm5sq}, we consider the unit disk containing $m^2$
identical disks arranged in a regular square-lattice packing.  Given
the non-overlapping condition, the maximal possible disk radius is
$R_{\mathrm{max}} = \frac{L}{\sqrt{2}(m-1)+1 }$, where $L = 1$ is the
outer boundary radius.  Keeping the obstacle centers fixed, we vary
their radii as $R = R_{\mathrm{max}} \times [0.01, 0.1, 0.2, 0.3, 0.5,
0.7, 0.9, 0.99, 1.0]$, to systematically investigate the disorder
influence on both the first-reaction time and the correlation
coefficient.  The reactive target is the central disk, whereas the
other disks are inert reflecting obstacles.  To interpret the overall
effect of the inert obstacles, we introduce an equivalent annulus
domain, whose effective outer radius $L_{\mathrm{eff}}$ is determined
by imposing the same area condition: $\pi (L_{\mathrm{eff}}^2 - R^2) =
\pi (L^2 - m^2 R^2)$, i.e.,
\begin{equation}
L_{\mathrm{eff}} = \sqrt{L^2 - (m^2-1) R^2}.  
\label{eq:annulus_app}
\end{equation}
Figure \ref{fig:T5} compares the Monte Carlo results for mean FRTs
obtained for $m=5$ with the annulus approximation (colored dashed
lines) from Eq. (\ref{eq:tau1_annulus}):
\begin{equation}
\langle \tau \rangle (\circ) = \frac{L_{\mathrm{eff}}^4 \log(L_{\mathrm{eff}}/R)}{2D (L_{\mathrm{eff}}^2 - R^2)} 
- \frac{3L_{\mathrm{eff}}^2 - R^2}{8D} + \frac{L_{\mathrm{eff}}^2 - R^2}{2q D R} . 
\label{eq:tau1_annulus_circ}
\end{equation}
The black solid curve corresponds to the perfect absorption limit
$q\to\infty$, i.e., the classical first-passage time.  When the disk
radius $R$ is small, the obstacles almost do not affect the search
process, except for reducing the space available for diffusion, and
the approximation agrees well with the simulations.  Moreover, as the
target (the central disk) has the same radius $R$, the mean FRT is
large (we recall that it diverges as $1/(qR)$ as $R\to0$). In turn, as
$R$ increases, the mean FRT decreases, until reaching its minimum. The
presence of this minimum is a peculiar feature of the considered
arrangement of obstacles.  In fact, in the limit case $R =
R_{\mathrm{max}}$, as starting points are uniform in the domain, some
of the particles cannot reach the reactive target, thus leading to
infinite FRT.  As this effect does not exist for the annulus domain,
noticeable deviations appear at larger radii.

It is worth noting that the annulus approximation was introduced
solely to illustrate the effect of the excluded volume in diluted
configurations of obstacles.  This basic approximation is not expected
to control connectivity and narrow-passage effects that dominate
first-passage and reaction statistics, as witnessed by
Fig. \ref{fig:m5_2d} (see deviations at large $R$).  In this light, it
was unexpected that such a crude approximation was good enough to
accurately reproduce a significant decrease of the mean FRT time as
$R/R_{\rm max}$ increases up to $0.4$ (that corresponds to the
occupation fraction of around $10\%$).

The correlation coefficient exhibits a similar trend. The equivalent
annulus approximation from Eq. (\ref{eq:Cq_annulus_circ}) remains
accurate for $R/R_{\mathrm{max}} < 0.3$ across all selected
reactivities.  In the high-reactivity limit $q \to \infty$, the power
law $C_q(\circ) \propto 1/q$ is recovered, as seen from the separation
between the blue ($q=10^3$) and orange ($q=10^2$) dashed curves shown
in Fig. \ref{fig:Cq5}.  As expected, when $q\to0$, one has $C_q(\circ)
\to 1$.  In the limit $R \to R_{\mathrm{max}}$, $C_q(\circ)$ shows a
sharp decrease.

\begin{figure}[t!]
  \centering
  \begin{subfigure}[b]{0.48\textwidth}
\includegraphics[width=0.99\linewidth]{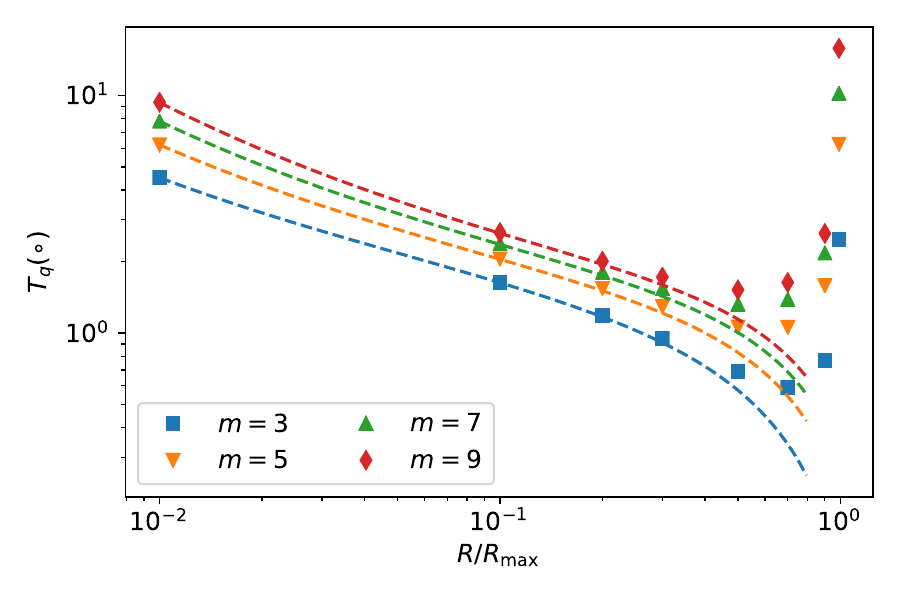}    \caption{Mean FRT}
    \label{fig:Tk2}
  \end{subfigure}
  \hspace{1.0em} 
  \begin{subfigure}[b]{0.48\textwidth}
\includegraphics[width=0.99\linewidth]{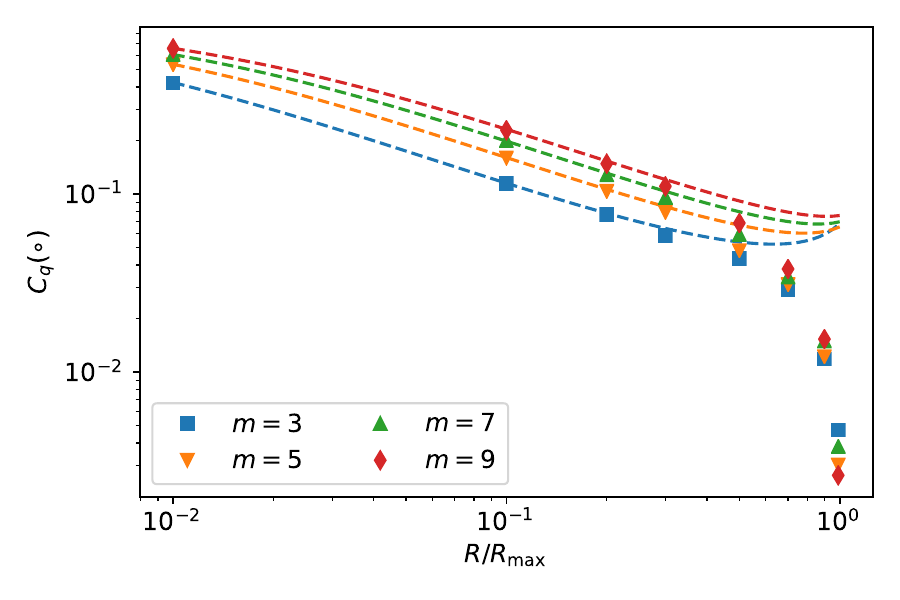}    \caption{Correlation coefficient}
    \label{fig:Cqk2}
  \end{subfigure}
\caption{
Role of the obstacle-lattice density $m$ at fixed reactivity $q=10^2$.
Numerical results are shown for: $m=3$ (squares), $m=5$ (downward
triangles), $m=7$ (upward triangles), and $m=9$ (diamonds).  We used
Monte Carlo simulations with $N=10^6$ particles that are initially
uniformly distributed inside the domain shown in Fig. \ref{fig:dm5sq}.
\textbf{(a)} Mean FRT as a function of the scaled disk radius $R /
R_{\mathrm{max}}$. Colored dashed curves show the annulus
approximation (\ref{eq:tau1_annulus_circ}) for each $m$.
\textbf{(b)} Correlation coefficient $C_q(\circ)$ as a function of the
scaled disk radius. Colored dashed curves show the annulus
approximation (\ref{eq:Cq_annulus_circ}), in which $L$ is replaced by
$L_{\mathrm{eff}}$ from Eq. (\ref{eq:annulus_app}).  }
\label{fig:k2_2d}
\end{figure}

We also investigate the effect of obstacle density by changing $m$
value. Note that the maximal radius $R_{\mathrm{max}}$ itself depends
on $m$ that also affects the horizontal axis in the figures, which is
the scaled radius $R / R_{\mathrm{max}}$.  As shown in
Fig. \ref{fig:Tk2}, the annulus approximation works if $R /
R_{\mathrm{max}} < 0.3$, and then the mean FRT increases due to
geometrical constraints. A larger density of obstacles (i.e., a larger
$m$) results in a larger mean FRT. Although the area occupation
fraction converges to a constant for a given ratio $R /
R_{\mathrm{max}}$ as $m \to \infty$ (see Eq. (\ref{eq:annulus_app})),
the mean FRT diverges according to Eq. (\ref{eq:tau1_annulus_circ})
since $R \lesssim R_{\mathrm{max}} \to 0$ as $m\to\infty$.
Consistently, the correlation coefficient also increases with $m$
(Fig. \ref{fig:Cqk2}).

\subsection{Randomly packed obstacles in two dimensions}
\label{sec:2drandom}

\begin{figure}[t!]
  \centering
  \begin{subfigure}[b]{0.48\textwidth}
\includegraphics[width=0.99\linewidth]{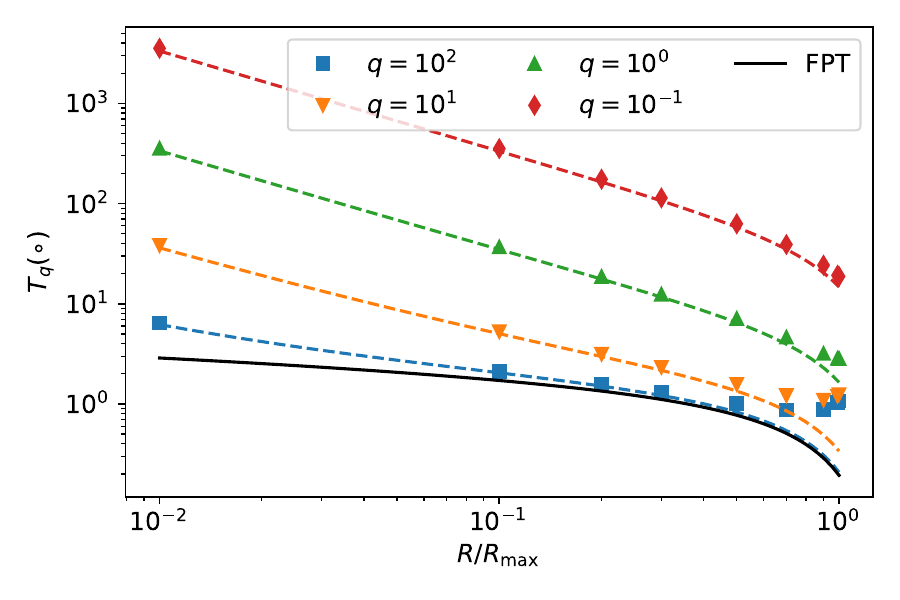}    \caption{Mean FRT}
    \label{fig:obs25T}
  \end{subfigure}
  \hspace{1.0em} 
  \begin{subfigure}[b]{0.48\textwidth}
\includegraphics[width=0.99\linewidth]{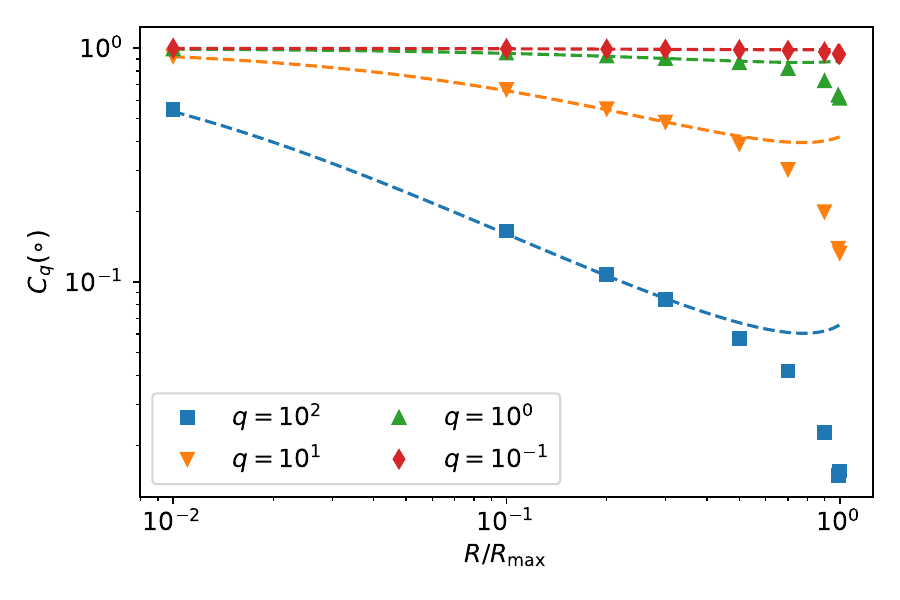}  \caption{Correlation coefficient}
    \label{fig:obs25C}
  \end{subfigure}
\caption{
Mean FRT and correlation coefficient for a random arrangment of
obstacles (the domain shown in Fig. \ref{fig:dm5r}).  Symbols
correspond to different reactivities: $q=10^2$ (squares), $q=10^1$
(downward triangles), $q=10^0$ (upward triangles), and $q=10^{-1}$
(diamonds).  Numerical results obtained by Monte Carlo simulations
with $N=10^6$ particles that are initially uniformly distributed
inside the domain.
\textbf{(a)} Mean FRT as a function of the scaled disk radius
$R/R_{\mathrm{max}}$. The black solid line represents the mean
first-passage time (i.e., $q\to\infty$), and colored dashed lines show
the annulus approximation (\ref{eq:tau1_annulus_circ}).  \textbf{(b)}
Correlation coefficient $C_q(\circ)$ as a function of the scaled disk
radius. Colored dashed lines show the annulus approximation
(\ref{eq:Cq_annulus_circ}), in which $L$ is replaced by
$L_{\mathrm{eff}}$ from Eq. (\ref{eq:annulus_app}).  }
\label{fig:obs_2d}
\end{figure}

In addition to the regular square-lattice packing, we examine another
configuration in which the obstacles are packed randomly.  To compare
with previous results for $m=5$, we consider the configuration shown
in Fig. \ref{fig:dm5r} (see Appendix \ref{app:random25} for
coordinates), with one reactive target at the center and 24 obstacles
packed randomly.  The maximal radius is determined by the 50\% area
occupation fraction. We consider the mean FRT (Fig. \ref{fig:obs25T})
and the correlation coefficient (Fig. \ref{fig:obs25C}) as before,
where the annulus approximations (\ref{eq:tau1_annulus_circ}) and
(\ref{eq:Cq_annulus_circ}) are also utilized with the effective radius
given by Eq. (\ref{eq:annulus_app}).  Compared with the regular
packing, the annulus approximation also works when $R /
R_{\mathrm{max}} < 0.3$ for both mean FRT and correlation
coefficient. The main difference arises in the limit $R \to
R_{\mathrm{max}}$, in which the mean FRT is finite, and the
correlation coefficient remains positive because the target remains
accessible from any point in the domain (i.e., there is no geometric
isolation of the target).

\subsection{Regularly packed obstacles in three dimensions}
\label{sec:3dregular}

\begin{figure}[t!]
  \centering
  \begin{subfigure}[b]{0.48\textwidth}
\includegraphics[width=0.99\linewidth]{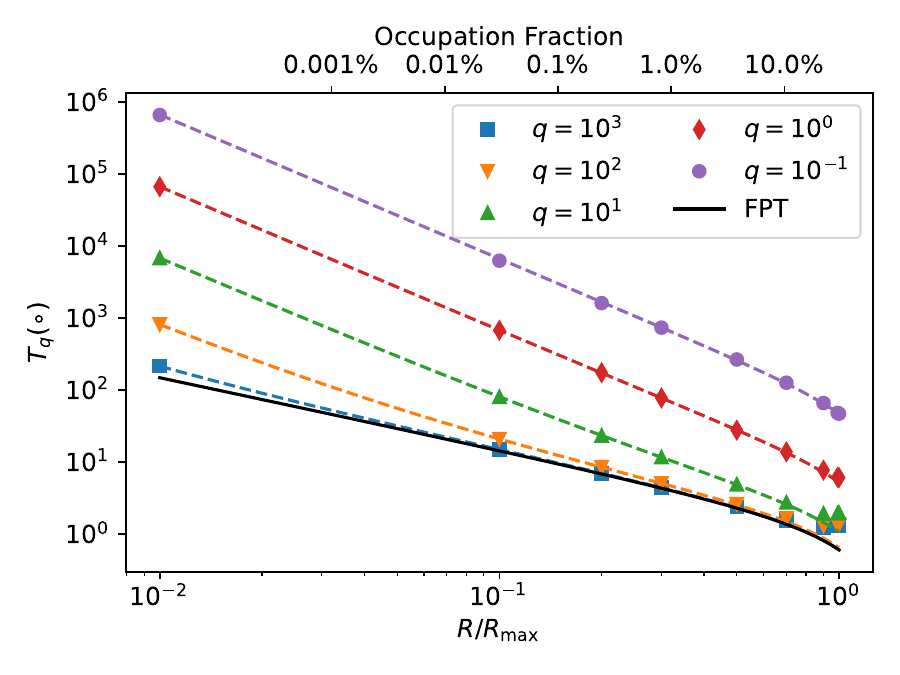}    \caption{Mean FRT}
    \label{fig:Td3_m3}
  \end{subfigure}
  \hspace{1.0em} 
  \begin{subfigure}[b]{0.48\textwidth}
\includegraphics[width=0.99\linewidth]{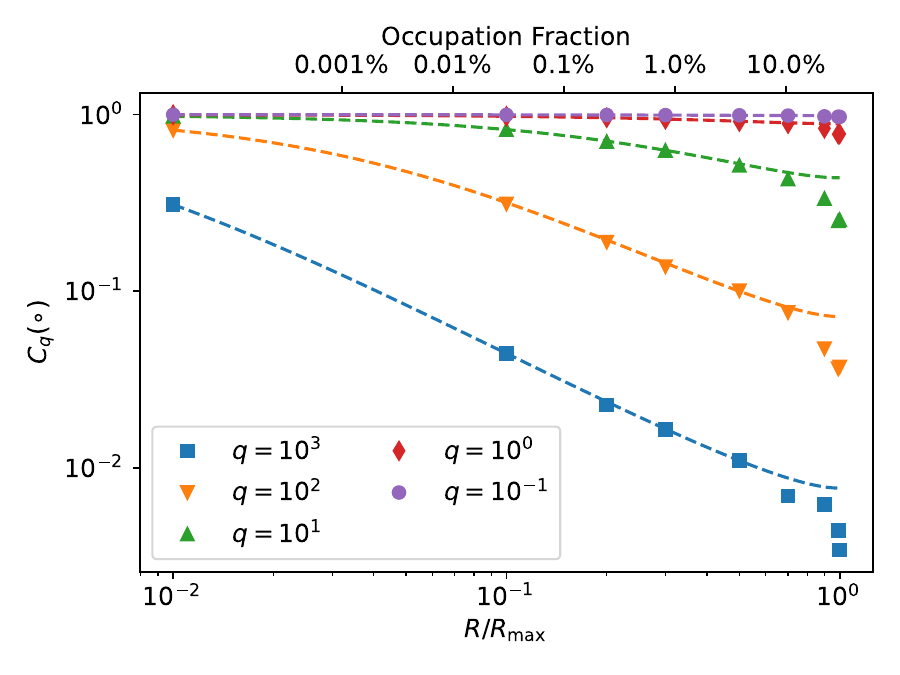}   \caption{Correlation coefficient}
    \label{fig:Cqd3_m3}
  \end{subfigure}
\caption{
Mean FRT and correlation coefficient for a regular arrangment of
obstacles inside the ball (the domain shown in Fig. \ref{fig:dm3d3}).
Symbols correspond to different reactivities: $q=10^3$ (squares),
$q=10^2$ (downward triangles), $q=10^1$ (upward triangles), $10^0$
(diamonds), and $q=10^{-1}$ (dots).  Numerical results obtained by
Monte Carlo simulations with $N=10^4$ particles that are initially
uniformly distributed inside the domain.
\textbf{(a)} Mean FRT as a
function of the scaled ball radius $R/R_{\mathrm{max}}$. The black
solid line represents the mean first-passage time (i.e.,
$q\to\infty$), and colored dashed lines show the shell approximation
(\ref{eq:tau1_shell_circ}).  \textbf{(b)} Correlation coefficient
$C_q(\circ)$ as a function of the scaled ball radius. Colored dashed
lines show the shell approximation (\ref{eq:Cq_shell_circ}), in which
$L$ is replaced by $L_{\mathrm{eff}}$ from Eq. (\ref{eq:shell_app}).
}
\label{fig:domain_3d}
\end{figure}

Beyond two-dimensional configurations, we explore the effect of
three-dimensional geometrical constraints via inert obstacles arranged
on a regular cubic lattice inside the unit ball, as illustrated by
Fig. \ref{fig:dm3d3}.  Again, given the non-overlapping condition, the
maximal radius of each obstacle is $R_{\mathrm{max}} =
L/(\sqrt{3}(m-1)+1)$, where $L=1$ is the outer boundary
radius. Keeping the obstacle centers fixed, we vary their radii for
numerical simulations.  The overall effect of the obstacles is
interpreted through an equivalent spherical shell domain, whose
effective outer radius $L_{\mathrm{eff}}$ is determined by imposing
the same volume condition, $\frac{4\pi}{3} (L_{\mathrm{eff}}^3 - R^3)
= \frac{4\pi}{3} (L^3 - m^3 R^3)$, i.e.,
\begin{equation}
L_{\mathrm{eff}} = \sqrt[3]{L^3 - (m^3-1) R^3}.
\label{eq:shell_app}
\end{equation}
Figure \ref{fig:Td3_m3} compares the Monte Carlo results at $m=3$ for
the mean FRT with this shell approximation (\ref{eq:tau1_shell_circ})
(colored dashed lines). The black solid curve corresponds to the
perfect absorption limit $q \to \infty$.  Unlike the two-dimensional
case, the approximation agrees well with simulations until a larger
value of $R/R_{\mathrm{max}}$; for instance, for $q=10^{-1}$, there is
still good agreement even at $R = R_{\mathrm{max}}$.  In this setting,
all particles can reach the reactive target within a finite FRT, since
the three-dimension constraint does not isolate the target like in the
two-dimensional case. The correlation coefficient
(Fig. \ref{fig:Cqd3_m3}) exhibits a similar trend. The equivalent
shell approximation (\ref{eq:Cq_shell_circ}) remains accurate for
$R/R_{\mathrm{max}} < 0.3$.  When the obstacle radius is close to the
maximum, $C_q$ exhibits a decrease, but remains strictly positive.

\section{Discussion and conclusion} 
\label{sec:discon}

In this study, we have systematically investigated the statistical
correlation between two fundamental quantities of diffusion-mediated
surface reactions: the first-reaction time $\tau$ and the acquired
boundary local time $\ell_\tau$. We developed a general theoretical
framework for accessing their joint probability density ${\mathcal
P}(\ell,t|\x_0)$.  For a broad class of diffusion-controlled
reactions, Eq. (\ref{eq:Pjoint}) decouples the effects of chemical
kinetics (incorporated through the probability density $\psi(\ell)$ of
the imposed threshold $\hat{\ell}$) and diffusive dynamics
(represented via the probability density $U(\ell,t|\x_0)$ of the
first-crossing time).  This multiplicative form is a consequence of
the assumed independence of these two ``ingredients''.  Despite an
apparent simplicity of this representation, the considered model is
fairly general; for instance, it includes the most common scenario of
a partially reactive region with a constant reactivity as a special
case.  The derived representation of the joint probability density
sheds light on correlations between $\tau$ and $\ell_\tau$ that are
characterized through the correlation coefficient $C_q$.
We derived explicit analytical expressions for $C_q$ in several basic
domains, including the ball, the disk, the concentric spherical shell,
and the concentric circular annulus. These exact results served as a
benchmark for validating our Monte Carlo simulation technique based on
the walk-on-spheres and the escape-from-a-layer algorithms.  As these
correlations are affected by the domain geometry, the analysis was
then extended to disordered environments with interior obstacles (2D
regular and random packings, as well as 3D regular packing).  By
employing concentric shell and annulus approximations, we examined the
influence of geometric disorder by varying the number of obstacles and
their radius.  An interesting finding in two-dimensional regular
configurations of obstacles is the emergence of a non-monotonic
behavior in the mean FRT, a direct consequence of geometrical
constraints.  In the limiting geometric configurations that completely
isolate a particle from the target, mean FRT diverges while the
cross-moment vanishes.  Across all other geometries, we confirmed the
universal limiting behaviors of the correlation: $C_q \to 1$ as $q \to
0$, indicating perfect correlation when reactions are rare and
boundary exploration is extensive; and $C_q \to 0$ as $q\to\infty$,
reflecting the decoupling due to instantaneous absorption upon the
first arrival.  An important theoretical perspective is the analysis
of the FRT, the BLT and their correlations for particles diffusing in
porous media, in which pores are separated by narrow passages (as in
the case of regularly arranged obstacles in two dimensions).
Developing refined analytical tools for such poorly connected
structures would be valuable.

This work opens several promising avenues for future research. For
instance, the framework can be generalized to domains with multiple,
disjoint reactive targets of different reactivities \cite{Galanti16a},
where the joint distribution of multiple BLTs and their
cross-correlations with the global FRT become relevant.  Also, our
findings on the strong dependence of $C_q$ on the geometry motivate a
detailed study of the optimization problem: how does the correlation
change with the shape, size, and spatial arrangement of reactive and
inert regions? Can one achieve the desired correlation by re-arranging
inert obstacles?  Furthermore, extending the analysis to
time-dependent reactivities $q(t)$ or spatially heterogeneous surface
properties would bridge the studied model closer to applications in
heterogeneous catalysis and dynamic biological interfaces.  Exploring
these directions will enhance our understanding of diffusion-mediated
surface reactions and provide refined tools for designing and
controlling processes in complex environments \cite{Dorsaz10,
Galanti16b} including porous media and cellular environments.

\section*{Acknowledgements}
D.S.G. acknowledges the Alexander von Humboldt Foundation for support
within a Bessel Prize award.

\section*{Data Availability Statement}
The data that support the findings of this study are available from
the corresponding author upon reasonable request.

\appendix
\section{Calculations}

For completeness, we provide intermediate results used in the
calculation of the correlation coefficients for several basic domains,
including the mean, second moment, and variance of the FRT, as well as
the raw correlation between the FRT and the BLT. We recall that
$(\circ)$ refers to the uniform starting point $\x_0 \in \Omega$ with
a constant density $\rho(\x) =|\Omega|^{-1}$.

\subsection{Ball}
\label{app:sphere}

When the starting point is uniformly distributed inside a ball of
radius $R$ (i.e., $\rho(\x) = |\Omega|^{-1} = 3/ (4\pi R^3)$), one has
\begin{align}
\tilde{U}(\ell,p|\circ) &= \frac{1}{|\Omega|} e^{-\mu_0^{(p)} \ell} 
\int\limits_\Gamma \mathrm{d} \x \ V_0^{(p)} (\x) \int\limits_\Omega \mathrm{d} \x_0 \ V_0^{(p)} (\x_0) \nonumber \\
&= \frac{D \mu_0^{(p)}}{p |\Omega|} |\Gamma| e^{-\mu_0^{(p)} \ell} ,
\label{eq:Uuni}
\end{align}
where $\Gamma = \pa$, $\alpha = \sqrt{p/D}$, $\mu_0^{(p)} = \alpha
i_0^\prime(\alpha R) / i_0(\alpha R)$, $V_0^{(p)}(\x) = (1 /
\sqrt{|\Gamma|}) i_0(\alpha |\x|) / i_0(\alpha R) $, and we used the
identity 
\begin{equation*}
\int_\Omega \mathrm{d} \x_0 \ V_0^{(p)} (\x_0) = \frac{D \mu_0^{(p)}}{p} \int_\pa \md \x \ V_0^{(p)} (\x) .  
\end{equation*}
Also, one has
\begin{align}
\tilde{H}_q(p|\circ) &= \frac{1}{|\Omega|} \int\limits_\Omega \md \x_0 \ \tilde{H}_q (p|\x_0) = \int\limits_0^\infty \mathrm{d} \ell \, q e^{-q \ell} \tilde{U}(\ell,p|\circ) \nonumber \\
&= \frac{q D \mu_0^{(p)} |\Gamma|}{p |\Omega| (q + \mu_0^{(p)})} ,
\label{eq:Htilde}
\end{align}
from which one gets the moments: 
\begin{subequations}
\begin{align}
\lla \tau \rra (\circ) &= \frac{R^2}{15D} + \frac{R}{3Dq} , \label{eq:tau1_sphere_unif} \\
\lla \tau^2 \rra (\circ) &= \frac{2R^2}{315D^2q^2} (35 + 14 qR + 2q^2R^2) , \\
\mathrm{Var}\{\tau\} (\circ) &= \frac{R^2}{1575 D^2 q^2} (13 q^2 R^2 + 70qR + 175) .
\end{align}
\end{subequations}
From Eq. (\ref{eq:Uuni}), we have
\begin{equation}
\lla \mathcal{T}_\ell \rra (\circ) = \frac{R^2}{15D} + \frac{R\ell}{3D} .
\end{equation}
Thus the cross-moment reads 
\begin{align}
\lla \tau \ell_\tau \rra (\circ) &= \int\limits_0^\infty \md \ell \ \ell \ qe^{-q\ell} \llp \frac{R^2}{15D} + \frac{R\ell}{3D} \rrp \nonumber \\
&= \frac{R^2}{15 Dq} + \frac{2R}{3Dq^2} ,
\end{align}
and we get 
\begin{equation}
\lla \tau \ell_\tau \rra - \lla \tau \rra \lla \ell_\tau \rra = \frac{R}{3Dq^2} ,
\end{equation}
which is the same as Eq. (\ref{eq:cm_sphere}) for a fixed starting point. 

\subsection{Disk}
\label{app:disk}

Consider diffusing particles inside the disk of radius $R$, whose
boundary $\Gamma = \pa$ is partially reactive with $q>0$.  For the
uniform starting point, both Eqs. (\ref{eq:Uuni}) and
(\ref{eq:Htilde}) remain valid, with $\alpha = \sqrt{p/D}$,
$\mu_0^{(p)} = \alpha I_1(\alpha R) / I_0(\alpha R)$, $V_0^{(p)}(\x) =
(1 / \sqrt{|\Gamma|}) I_0(\alpha |\x|) / I_0(\alpha R) $, where
$I_n(z)$ is the modified Bessel function of the first kind. In turn,
when the starting point is fixed, one has
\begin{align}
\tilde{U}(\ell,p|\x_0) &= e^{-\mu_0^{(p)} \ell} V_0^{(p)}(\x_0) \int\limits_\Gamma \md\x \ V_0^{(p)}(\x) \nonumber \\
&= e^{-\mu_0^{(p)} \ell} \frac{I_0(\alpha |\x_0|)}{I_0(\alpha R)} ,
\end{align}
and thus 
\begin{align}
\tilde{H}_q (p|\x_0) &= \int\limits_0^\infty \mathrm{d} \ell \ q e^{-q\ell} \ \tilde{U}(\ell,p|\x_0) \nonumber \\
&= \frac{q I_0(\alpha |\x_0|)}{\alpha I_1(\alpha R) + q I_0(\alpha R)} .
\end{align}
The mean FRTs are 
\begin{subequations}
\begin{align}
\langle \tau \rangle (r) &= \frac{R^2 - r^2}{4D} + \frac{R}{2 D q} , \\
\langle \tau \rangle (\circ) 
&= \frac{R^2}{8D} + \frac{R}{2 D q} . 
\end{align}
\end{subequations}
Solving Eqs. (\ref{eq:taumrecurr}) for $m=2$ implies
\begin{subequations}
\begin{align}
\langle \tau^2 \rangle (r) &= \frac{R^2}{2 D^2 q^2} + \frac{3 R^3-2 r^2 R}{8 D^2 q} + \frac{r^4-4 r^2 R^2+3 R^4}{32 D^2} , \\
\langle \tau^2 \rangle (\circ) &= \frac{R^2}{2 D^2 q^2}+\frac{R^3}{4 D^2 q}+\frac{R^4}{24 D^2} , 
\end{align}
\end{subequations}
and thus the variances read
\begin{subequations}
\begin{align}
\mathrm{Var}\{\tau\}(r) &= \frac{q^2 \left(R^4-r^4\right)+4 q R^3+8 R^2}{32 D^2 q^2} , \\
\mathrm{Var}\{\tau\} (\circ) &= \frac{R^2 (q R (5 q R+24)+48)}{192 D^2 q^2} .
\end{align}
\end{subequations}
Moreover, we have
\begin{subequations}
\begin{align}
\langle \mathcal{T}_\ell \rangle (r) &= \frac{R^2 - r^2}{4D} + \frac{R \ell}{2 D},  \\
 \langle \mathcal{T}_\ell \rangle (\circ) &= \frac{R^2}{8 D} + \frac{R \ell}{2 D} . 
\end{align}
\end{subequations}
As a result, we get the cross-moments
\begin{subequations}
\begin{align}
\langle \tau \ell_\tau \rangle (r) &= \int\limits_0^\infty \md\ell \, \ell \ q e^{-q\ell} \langle \mathcal{T}_\ell \rangle (r) 
= \frac{q \left(R^2-r^2\right)+4 R}{4 D q^2} , \\
\langle \tau \ell_\tau \rangle (\circ) &= \int\limits_0^\infty \md\ell \, \ell \ q e^{-q\ell} \langle \mathcal{T}_\ell \rangle (\circ) 
= \frac{R (q R+8)}{8 D q^2} ,
\end{align}
\end{subequations}
so that 
\begin{equation}
\langle \tau \ell_\tau \rangle - \langle \tau \rangle \langle \ell_\tau \rangle = \frac{|\Omega|}{|\partial \Omega| Dq^2} 
= \frac{R}{2 D q^2} \end{equation}
for both cases of a fixed and uniformly distributed starting point. 
The correlation coefficient $C_q$ then follows as
\begin{subequations}
\begin{align}
C_q (r) &= \frac{1}{\sqrt{1 + qR/2 + (qR)^2 \left(1-(r/R)^4\right) / 8}} , \\
C_q (\circ) &= \frac{1}{\sqrt{1 + q R /2 + 5 (q R)^2 /48}} .
\label{eq:Cq_disk}
\end{align}
\end{subequations}

\subsection{Annulus}
\label{app:annulus}

Consider particles diffusing inside two-dimensional concentric
circular annulus of radii $R<L$, where the inner boundary $\Gamma$ (at
$r=R$) is partially reactive with $q>0$ and the outer one $\pa_N$ (at
$r=L$) is inert.  Solving Eqs. (\ref{eq:taumrecurr}) for $m=1$, we get
the mean FPTs:
\begin{subequations}
\begin{align}
\langle \tau \rangle (r) &= \frac{L^2}{2D} \log \llp \frac{r}{R} \rrp + \frac{R^2 - r^2}{4D} + \frac{L^2 - R^2}{2q D R} , \\
\langle \tau \rangle (\circ) &= \frac{L^4 \log(L/R)}{2D (L^2 - R^2)} - \frac{3L^2 - R^2}{8D} + \frac{L^2 - R^2}{2q D R} . \label{eq:tau1_annulus}
\end{align}
\end{subequations}
Solving Eqs. (\ref{eq:taumrecurr}) for $m=2$ yields 

\begin{widetext}
\begin{subequations}
\label{eq:tau2_annulus}
\begin{align}
\langle \tau^2 \rangle(r) &= \frac{1}{32 D^2 q^2 R^2} \Bigg\{ 
\begin{aligned}[t]
    & 4 L^4 (4-3 q R) -8 L^2 R \left(q r^2 (1-q R)+q R^2 (q R-3)+4 R\right) \\ &+R^2 \left(q^2 \left(r^4-4 r^2 R^2+3 R^4\right) +4 q \left(2 r^2 R-3 R^3\right)+16 R^2\right) \\
& + 4 q R L^2 \left[ 4 L^2 \log \left(\frac{L}{R}\right) \left(q R \log \left(\frac{r}{R}\right)+1\right)+ \log \left(\frac{r}{R}\right) \left(L^2 (4-3 q R) - 2 R \left(q r^2-q R^2+2 R\right)\right)\right] \Bigg\} ,
\end{aligned}
\label{eq:tau2_annulus_r} \\
\langle \tau^2 \rangle (\circ) &= \frac{1}{48 D^2 q^2 R^2 \left(L^2-R^2\right)} \Bigg\{ 
\begin{aligned}[t]
& \left(L^2-R^2\right) \left(L^4 \left(17 q^2 R^2-36 q R+24\right) - L^2 R^2 \left(13 q^2 R^2 - 48 q R + 48\right)+2 R^4 \left(q^2 R^2-6 q R+12\right)\right) \\
& - 12 L^4 q R \log \left(\frac{L}{R}\right) \left(L^2 (3 q R-4) - 2 L^2 q R \log \left(\frac{L}{R}\right)-2 R^2 (q R-2)\right) \Bigg\} . 
\end{aligned}
\label{eq:tau2_annulus_circ}
\end{align}
\end{subequations}
\end{widetext}

By replacing $1/q$ as $\ell$ in $\langle \tau \rangle$, one gets 
\begin{subequations}
\begin{align}
\langle \mathcal{T}_\ell \rangle (r) &= \frac{L^2}{2D} \log \llp \frac{r}{R} \rrp + \frac{R^2 - r^2}{4D} + \frac{L^2 - R^2}{2D R} \ell , \\
\langle \mathcal{T}_\ell \rangle (\circ) &= \frac{L^4 \log(L/R)}{2D (L^2 - R^2)} - \frac{3L^2 - R^2}{8D} + \frac{L^2 - R^2}{2D R} \ell .
\end{align}
\end{subequations}
The cross-moment reads
\begin{subequations}
\begin{align}
\langle \tau \ell_\tau \rangle (r) 
&= \frac{L^2 - R^2}{D q^2 R} + \frac{R^2-r^2}{4 D q} + \frac{L^2 \log(r/R)}{2 Dq } , \\
\langle \tau \ell_\tau \rangle (\circ) 
&= \frac{L^2 - R^2}{D q^2 R} + \frac{R^2- 3L^2}{8 D q} + \frac{L^2 \log(L/R)}{2 Dq (L^2-R^2) } ,
\end{align}
\end{subequations}
so that 
\begin{equation}
\langle \tau \ell_\tau \rangle - \langle \tau \rangle \langle \ell_\tau \rangle = \frac{|\Omega|}{|\partial \Omega| Dq^2} = \frac{L^2-R^2}{2 D q^2 R} 
\end{equation}
in both cases.  Combining these results, one determines $C_q(\circ)$
from Eq. (\ref{eq:cq}) for the annulus approximation used in
Fig. \ref{fig:Cq5}, \ref{fig:Cqk2}, and \ref{fig:obs25C}:

\begin{widetext}
\begin{align}
C_q (\circ) &= \frac{4 \sqrt{3} (L^2-R^2)}{\sqrt{ \begin{aligned}[t]
    & L^4 (q R (41 q R-72)+48) -2 L^2 R^2 (q R (17 q R-48)+48) +R^4 (q R (5 q R-24)+48) \\
    & +24 L^4 q R \log \left(\frac{L}{R}\right) \left(\frac{2 L^2 q R \left(L^2-2 R^2\right) \log \left(\frac{L}{R}\right)}{\left(L^2-R^2\right)^2}-3 q R+4\right) 
    \end{aligned} }} \ .
    \label{eq:Cq_annulus_circ}
\end{align}
\end{widetext}

\subsection{Spherical shell}
\label{app:shell}

For the three-dimensional concentric spherical shell of radii $R<L$
with the reactive inner boundary and the inert outer one, the mean
FRTs read
\begin{subequations}
\begin{align}
\langle \tau \rangle(r) &= \frac{L^3-R^3}{3 D q R^2} + \frac{2 L^3 (r-R)-r^3 R+r R^3}{6 D r R} , \\
\langle \tau \rangle (\circ) &= \frac{L^3-R^3}{3 D q R^2} + \frac{(L-R)^2 \left(5 L^3+6 L^2 R+3 L R^2+R^3\right)}{15 D R \left(L^2+L R+R^2\right)} , 
\label{eq:tau1_shell_circ}
\end{align}
\end{subequations}
where we used a uniform distribution $\rho(\x) = |\Omega|^{-1} =
3/(4\pi (L^3 - R^3))$. Then, one solves Eqs. (\ref{eq:taumrecurr})
with $m=2$ for the second moment of FRTs:

\begin{widetext}
\begin{subequations}
\label{eq:tau2_shell}
\begin{align}
\langle \tau^2 \rangle (r) &= \frac{1}{180 D^2 q^2 r R^4} \bigg\{ 
\begin{aligned}[t]
    & 40 L^6 (q R+1) \left(q r R-q R^2+r\right) -72 L^5 q R^2 \left(q r R-q R^2+r\right) \\
    & -20 L^3 R^2 \left(q^2 R \left(r^3-3 r^2 R+r R^2+R^3\right) + q \left(r^3-r R^2-2 R^3\right)+4 r R\right) \\
    & + r R^4 \left(q^2 \left(3 r^4-10 r^2 R^2+7 R^4\right) + 4 q \left(5 r^2 R-7 R^3\right)+40 R^2\right) \bigg\}, 
\end{aligned}
\label{eq:tau2_shell_a} \\
\langle \tau^2 \rangle (\circ) &= \frac{2}{315 D^2 q^2 R^4 \left(L^3-R^3\right)} \bigg\{  
\begin{aligned}[t]
    & 35 L^9 (q R+1)^2 - 126 L^8 q R^2 (q R+1) + 135 L^7 q^2 R^4 - 105 L^6 R^3  \\ 
    & - 63 L^5 q R^5 (q R-2) + 21 L^3 R^6 (q R (q R-4)+5) + R^9 (-2 q R (q R-7)-35) \bigg\} , 
\end{aligned}
\label{eq:tau2_shell_b}
\end{align}
\end{subequations}
\end{widetext}
thus determining the variances $\mathrm{Var} \{\tau\}(r)$ and
$\mathrm{Var} \{\tau\}(\circ)$.

Moreover, by replacing $1/q$ by $\ell$ in $\langle \tau \rangle$, one has mean FCTs:
\begin{subequations}
\begin{align}
\langle \mathcal{T}_\ell \rangle (r) &= \frac{(L^3-R^3)\ell}{3 D R^2} + \frac{2 L^3 (r-R)-r^3 R+r R^3}{6 D r R} , \\
\langle \mathcal{T}_\ell \rangle (\circ) &= \frac{(L^3-R^3) \ell}{3 D R^2} \nonumber \\
&+ \frac{(L-R)^2 \left(5 L^3+6 L^2 R+3 L R^2+R^3\right)}{15 D R \left(L^2+L R+R^2\right)} , 
\end{align}
\end{subequations}
so that the cross-moments read
\begin{subequations}
\begin{align}
\langle \tau \ell_\tau \rangle(r) 
=& \frac{2 \left(L^3-R^3\right)}{3 D q^2 R^2} + \frac{2 L^3 (r-R)-r^3 R+r R^3}{6 D q r R} , \\ 
\langle \tau \ell_\tau \rangle (\circ) 
=& \frac{2 \left(L^3-R^3\right)}{3 D q^2 R^2} \nonumber \\ &+ \frac{(L-R)^2 \left(5 L^3+6 L^2 R+3 L R^2+R^3\right)}{15 D q R \left(L^2+L R+R^2\right)} , 
\end{align}
\end{subequations}
from which  
\begin{equation} \langle \tau \ell_\tau \rangle - \langle \tau \rangle \langle \ell_\tau \rangle 
= \frac{|\Omega|}{|\partial \Omega| Dq^2} = \frac{L^3-R^3}{3 D q^2 R^2}  
\end{equation}
in both cases.  Combining these results, one determines $C_q(\circ)$
from Eq. (\ref{eq:cq}) for the shell approximation used in
Fig. \ref{fig:Cqd3_m3}:

\begin{widetext}
\begin{align}
C_q (\circ) &= \frac{5 \sqrt{7} \left(L^3-R^3\right)^2}{\sqrt{ \begin{aligned}[t]
    & \bigg\{ 10 \left(L^3-R^3\right) \big(35 L^9 (q R+1)^2 -126 L^8 q R^2 (q R+1)+135 L^7 q^2 R^4 -105 L^6 R^3 \\
    & -63 L^5 q R^5 (q R-2) +21 L^3 R^6 (q R (q R-4)+5) +R^9 (-2 q R (q R-7)-35)\big) \\
    & -7 \left(5 L^6 (q R+1)-9 L^5 q R^2 +5 L^3 R^3 (q R-2)+R^6 (5-q R)\right)^2 \bigg\}
    \end{aligned} }} \ . 
    \label{eq:Cq_shell_circ}
\end{align}
\end{widetext}

\section{Random packing of obstacles in two dimensions} 
\label{app:random25}

As shown in Fig. \ref{fig:dm5r}, the area occupation ratio is 50\%,
including the reactive target and inert obstacles. The maximal radius
is $1/(5\sqrt{2}) \approx 0.141421$.  The center of reactive target is
located at
$( 0.0 , 0.0 )$, while the coordinates of each obstacle's center read
$( -0.07068 , 0.3746 )$, 
$( 0.1703 , -0.786 )$, 
$( 0.4497 , -0.7037 )$, 
$( -0.343 , 0.4786 )$, 
$( -0.5462 , -0.02561 )$, 
$( 0.3037 , 0.3557 )$, 
$( -0.6671 , -0.5119 )$, 
$( -0.3806 , -0.6381 )$, 
$( -0.1114 , -0.4825 )$, 
$( 0.4061 , 0.05633 )$, 
$( 0.3541 , -0.2613 )$, 
$( 0.7377 , 0.01179 )$, 
$( -0.6381 , 0.331 )$, 
$( -0.2481 , -0.2096 )$, 
$( 0.4168 , 0.6535 )$, 
$( -0.8433 , 0.1079 )$, 
$( 0.07206 , 0.7421 )$, 
$( -0.7852 , -0.2465 )$, 
$( 0.6535 , 0.3383 )$, 
$( -0.2848 , 0.1658 )$, 
$( -0.1314 , -0.8289 )$, 
$( 0.6499 , -0.3344 )$, 
$( -0.2291 , 0.7764 )$, 
$( 0.1874 , -0.4939 )$.

\end{document}